\documentclass[journal]{IEEEtran}
\usepackage{epsfig,cite}
\usepackage{graphicx}
\usepackage{color}
\usepackage{amssymb,amsmath,mathtools, mathrsfs, dsfont}
\usepackage{enumitem}
\usepackage{graphicx}
\usepackage{epstopdf}
\definecolor{blue}{rgb}{0,0,1}
\definecolor{darkgreen}{rgb}{0,.5,0}
\definecolor{darkred}{rgb}{.5,0,0}

\newtheorem{theorem}{Theorem}

\newtheorem{lemma}{Lemma}
\newtheorem{rem}{Remark}

\newtheorem{definition}{Definition}
\newtheorem{cor}{Corollary}

\def\levy{L\'evy }

\renewcommand{\vec}[1]{\mathbf{#1}}

\newcommand{\realSet}{\mathcal{R}}
\newcommand{\intSet}{\mathcal{N}}

\newcommand{\styp}{\mathcal{T}_{\epsilon}^{(K)}}
\newcommand{\var}{\mathrm{Var}}

\def\ith{\text{$i^{\text{th}}$ }}

\def\LevyDist{{\mathscr{L}}}
\def\GumbDist{{\mathscr{G}}}
\def\NormDist{{\mathscr{N}}}
\def\Pr{{\mathrm{Pr}}}

\DeclareMathOperator\erfc{erfc}
\DeclareMathOperator\Ei{Ei}
\DeclareMathOperator\erfcinv{erfcinv}
\DeclareMathOperator\erfinv{erfinv}
\newcommand*{\dprime}{^{\prime\prime}\mkern-1.2mu}

\IEEEoverridecommandlockouts

\newcounter{MYtempeqncnt}

\begin{document}
\date{}

\title{Capacity Limits of Diffusion-Based \\Molecular Timing Channels}

\author{ \IEEEauthorblockN{Nariman~Farsad,~\IEEEmembership{Member,~IEEE,}
		Yonathan~Murin,~\IEEEmembership{Member,~IEEE,}\\
		Andrew~Eckford,~\IEEEmembership{Senior Member,~IEEE,} 
		and Andrea Goldsmith,~\IEEEmembership{Fellow,~IEEE}}
	
	\thanks{Nariman Farsad, Yonathan Murin, and Andrea Goldsmith are with the Department of Electrical Engineering, Stanford University, Stanford, CA, 94305 USA. Andrew Eckford is with the Department of Electrical Engineering and Computer Science, York University, Toronto, ON, M3J 1P3 Canada.}
	\thanks{Parts of this work were presented at the IEEE International Symposium on Information Theory (ISIT), July 2016, Barcelona, Spain, \cite{isit16}, and at the 50$^{\text{th}}$ Asilomar Conference on Signals, Systems, and Computers, Pacific Grove, CA,~\cite{asilomar16}.}
	\thanks{This research was supported in part by the NSF Center for Science of Information (CSoI) under grant CCF-0939370, and the NSERC Postdoctoral Fellowship fund PDF-471342-2015.}
	\thanks{Corresponding email: nfarsad@stanford.edu}
}
\maketitle

\vspace{-.7cm}
\begin{abstract}
	This work introduces capacity limits for molecular timing (MT) channels, where information is modulated in the release timing of small information particles, and decoded from the time of arrivals at the receiver. It is shown that the random time of arrival can be represented as an additive noise channel, and for the diffusion-based MT (DBMT) channel this noise is distributed according to the \levy distribution. Lower and upper bounds on the capacity of the DBMT channel are derived for the case where the delay associated with the propagation of the information particles in the channel is finite, namely, when the information particles dissipate after a finite time interval. For the case where a single particle is released per channel use, these bounds are shown to be tight. When the transmitter {\em simultaneously} releases a {\em large} number of particles, the detector at the receiver may not be able to precisely detect the arrival time of {\em all} the particles.	Therefore, two alternative models are considered: detection based on the particle that arrives first, or detection based on the {\em average} arrival times. Lower and upper bounds on the capacities of these two models are derived, and the lower bound also provides a lower bound for the capacity of the DBMT channel. It is shown that by controlling the lifetime of the information particles, the capacity can increase {\em poly-logarithmically} with the number of released particles. As each particle takes a random {\em independent} path, this diversity of paths is analogous to receiver diversity and can be used to considerably increase the achievable data rates.
\end{abstract}

\vspace{-.3cm}

\section{Introduction}

Molecular communication is an emerging field where small particles such as molecules are used to transfer information \cite{far16ST}. Information can be modulated on different properties of these particles such as their concentration \cite{kur12}, the type \cite{kim13}, the number \cite{far15TNANO}, or the time of release \cite{far15GLOBCOM}. Moreover, different techniques can be used to transfer the particles from the transmitter to the receiver including: diffusion \cite{mah10}, active transport \cite{far12NANO}, bacteria \cite{lio12}, and flow \cite{bic13}. To show the feasibility of molecular communication, in recent years a number of experimental systems have been developed that are capable of transmitting short messages at low bit rates \cite{far13,far14INFOCOM,lee2015_infocom}.

Despite all these advancements, there are still many open problems in the field, especially from an information theoretic perspective. For example, the fundamental channel capacity limits of many different molecular communication systems are still unknown \cite{far16ST}, particularly those with
indistinguishable molecules \cite{rose14}. 
Some of the challenge here is due to differences in the nature of conventional versus molecular communication systems, which must be considered in the capacity definition for the latter type of system. For example, in traditional electromagnetic communication the capacity does not depend on the symbol duration, and hence capacity can be defined in bits per channel use or in bits per second for a fixed symbol duration \cite[Ch. 8.1]{gallager-book}. In molecular communication, however, the symbol duration affects diffusion-based propagation and hence the channel, thus effecting the capacity. The notion of capacity per channel use depends on the symbol duration over which the channel is used, in contrast to electromagnetic communication. 

The first engineered molecular communication systems used concentration-modulation, whereby information is modulated based on the concentration of the released particles. In \cite{pie13}, a lower bound for the capacity of concentration-modulated channels in gaseous environments was presented. An achievable information rate, and a capacity expression for the time-slotted concentration-modulated molecular communication channel, were developed in \cite{ein2011} and \cite{ein11ITW}. In these channels, at the beginning of each time slot different concentrations of information particles are released by the transmitter to represent different symbols. The receiver uses the perceived concentration during the same time slot to detect the symbol, while it is assumed that the information particles which did not arrive within this time slot are destroyed.
The capacity in these works was defined in terms of the mutual information between the number of particles released and the number of particles that arrived during a symbol duration. In \cite{ata13}, the optimal input distribution for this channel was presented. Since the information particles may degrade over time, a capacity expression for concentration-modulated communication with degradable particles was  developed in \cite{nak12}. A Markov chain channel model for active transport molecular communication, where information particles are actively transported using molecular motors instead of diffusion, was derived in \cite{far14TSP}, which also presented the capacity of these channels.

\begin{figure}
	\begin{center}
		\includegraphics[width=1\columnwidth,keepaspectratio]{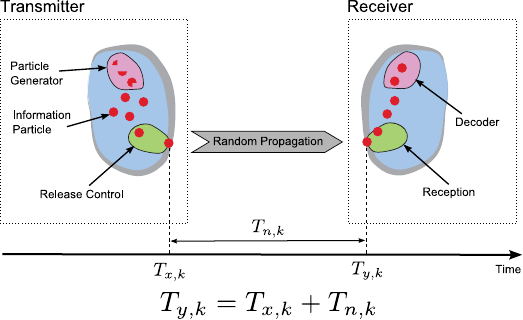}
	\end{center}
	\vspace{-0.3cm}
	\caption{\label{fig:diffuseMolComm} Diffusion-based molecular communication timing channel. $T_{x,k}$ denotes the release time, $T_{n,k}$ denotes the random propagation time, and $T_{y,k}$ denotes the arrival time.}
	\vspace{-0.25cm}
\end{figure}

In this work, we consider molecular communication systems where information is modulated on the {\em time of release of the information particles}, which is similar to pulse position-modulation \cite{shiu99}. 
Encoding information in the timing of transmission is not a new idea. For instance, \cite{borst99} used this approach to describe communication in the brain at the synaptic cleft, where two chemical synapses communicate over a chemical channel, and \cite{krishnaswamy13} used this model to study bacterial communication over a microfluidic chip. We refer the reader to \cite[Sec. II]{rose:InscribedPart1} for a detailed discussion about applications of timing-based communications in biology.
A common assumption, which is accurate for many sensors, is that the particle is detected and is removed from the environment as part of the detection process. Thus, the random delay until the particle first arrives at the receiver can be represented as an additive noise term. For example, for diffusion-based channels, the random first time of arrival is L\'evy-distributed \cite{book-firstPass, yilmaz20143dChannelCF}. Fig.~\ref{fig:diffuseMolComm} depicts these channels.

Note that although there are similarities between the timing channel considered in this work and the timing channels considered in \cite{ana96}, which studied the transmission of bits through queues, the problem formulation and the noise models are fundamentally different. In \cite{ana96}, the channel output (i.e. arrival times) from consecutive channel uses are {\em ordered}. This means that the first arrival time corresponds to the first channel use, the second arrival corresponds to second channel use, and so on. For molecular channels with {\em indistinguishable} particles, the information particle released during the first channel use may arrive after the information particle released in the second channel use. Therefore, the order of the transmitted information particles {\em may not be preserved} at the receiver as was observed in \cite{rose:InscribedPart1, rose:InscribedPart2}. 
Regarding the differences in the noise models we note that in \cite{ana96} the random delay is governed by the queue’s service distribution, while in molecular communication the random delay is associated with the transport of information particles in molecular channels.

Some of the previous works on molecular timing channels focused on the additive inverse Gaussian noise (AIGN) channel, which features a positive drift from the transmitter to the receiver. In this case, the first time of arrival over a one-dimensional space follows the inverse Gaussian distribution \cite{chhikara-folks}, giving the channel its name. 
In \cite{sri12}, upper and lower bounds on the maximal mutual information between the AIGN channel input and output, per channel use, were presented under the assumption that the average particle arrival time is constrained (i.e. is less than a constant). We denote this maximal mutual information as the {\em capacity per channel use}.
The same constraint was used in \cite{cha12}, which presented a different set of bounds on the capacity per channel use for the AIGN channel. A different constraint, which limits the maximum particle arrival time, was considered in \cite{eck12}, where an upper bound on the capacity per channel use was derived. Finally, \cite{li14} tightened the bounds derived in \cite{sri12} and \cite{cha12}, and characterized the capacity-achieving input distribution which can be used to accurately evaluate the capacity per channel use for the AIGN channel.

One of the main unresolved issues in these previous works is the problem of ordering, namely, information particles may arrive in an order different that the order they were released. 
Thus, it is not clear from \cite{sri12, cha12,eck12,li14} how information can be transmitted sequentially, and the associated capacity in {\em bits per second}. A partial answer for this question was provided in \cite{nanoComNet} that studied {\em time-slotted} transmission over MT channels {\em without} drift. Yet, the work \cite{nanoComNet} only provides a (sub-optimal) transmission scheme, leaving open the question of capacity for this channel.
To deal with the challenge of characterizing the fundamental capacity of diffusion-based molecular timing (DBMT) channels, in this work we make two assumptions. First, we assume that there is a finite time interval called the {\em symbol interval} over which the transmitter can encode its messages by choosing a specific time in this interval to release particles.
Second, we assume that the information particles have a finite lifespan, which we call the {\em particle's lifetime}. The underlying assumption is that the particles are dissipated immediately after this time interval.
We note that this assumption can be incorporated into a system by using enzymes or other chemicals that degrade the particles \cite{noe14,guo15}; as long as the particle's lifetime is less than infinity, our results and analysis hold. 
Using these assumptions, a {\em single channel use interval} is the sum of the symbol interval and the particle's lifetime, and information particles arrive during the same channel use in which they were released, or they dissipate over this interval and hence never arrive. 
 
The above assumptions enforce an ordering where particles arrive in the same order in which they are transmitted, resulting in identical and independent consecutive channel uses. We refer to this channel as the {\em molecular timing} (MT) channel, and note that it can be used with any propagation mechanism as long as the particles follow independent paths, and have a finite lifetime and symbol interval. 
Using this formulation, we define the capacity of the MT channel in bits per second. We then apply this definition to the DBMT channel, where the particles follow a Brownian path from the transmitter to the receiver, and derive an upper and a lower bound on the capacity in bits per second for the case where a {\em single} particle is transmitted per channel use. Through numerical evaluations we demonstrate that these bounds can be tight. 

When the transmitter {\em simultaneously} releases multiple particles, we consider three different receivers, and this leads to three {\em different} channel models. 
First, we consider a receiver that detects the {\em arrival time of each particle} and derive an expression for the capacity of the corresponding channel model.  
Since evaluating this capacity expression analytically seems intractable, we derive an upper bound that scales linearly with the number of released particles. 
Second, we consider a receiver that detects the time of the {\em first arrival} (FA). We demonstrate that the resulting system can be modeled by an additive noise channel, and for a large number of particles released, the noise is Gumbel distributed. We then derive the asymptotic lower and upper bounds on the capacity of this channel. 
Finally, we consider a system where the receiver detects the {\em average} arrival time of particles. 
A possible method to estimate this average arrival time is via measuring the number of particles arriving during sampled time intervals. 
We show that this system can be modeled as an additive noise channel, where for a large number of released particles, the noise is Gaussian distributed. Asymptotic lower and upper bounds on the capacity of this channel are presented. 
We emphasize that the lower bounds on capacity of the systems with the FA and the average detectors also serve as a lower bound on the capacity of the system that can detect all the arrival times of particles, i.e., these bounds also serve as a lower bound on the capacity of the DBMT channel without any constraints on the receiver.
Moreover, we show that by controlling the particles' lifetime, the capacity of the channels corresponding to these receivers increases at least {\em poly-logarithmically} with the number of particles, and that the average detector achieves higher information rates than the first arrival detector. In these systems, the increase in capacity is reminiscent of the capacity gains through receiver diversity in electromagnetic communication as each particle takes a random independent path from the transmitter to the receiver. 

The rest of this paper is organized as follows. The channel models for the MT and DBMT channels are presented in Section \ref{sec:model}. The capacity of the single-particle DBMT channel is studied in Section \ref{sec:singlePart}. The results are extended to the case of multiple particles in Sections \ref{sec:multiPart} and \ref{sec:FA_AVG}. The numerical evaluations are presented in Section \ref{sec:NumEvaluation}, and concluding remarks are provided in Section \ref{sec:concl}.

\section{System Model and Problem Formulation}
\label{sec:model}

\subsection{Notation}

We denote the set of real numbers by $\realSet$, the set of positive real numbers by $\realSet^{+}$, the set of positive natural numbers by $\intSet$, and the empty set by $\phi$. Other than these sets, we denote sets with calligraphic letters, e.g., $\mathcal{J}$, where $|\mathcal{J}|$ denotes the cardinality of the set $\mathcal{J}$.
We denote RVs with upper case letters, $X$, $Y$, $T$, and $\Theta$, their realizations with the corresponding lower case letters, e.g., $x$, $y$, and vectors with boldface letters, e.g., $\vec{X}, \vec{Y}$. The $i^{\text{th}}$ element of a vector $\vec{X}$ is denoted by $\vec{X}[i]$. All other upper case letters such as $D$, $K$, and $M$ are used to represent constants. 
We use $f_{Y}(y)$ to denote the probability density function (PDF) of a continuous RV $Y$ on $\realSet$, $f_{Y|X}(y|x)$ to denote the conditional PDF of $Y$ given $X$, and  $F_{Y}(y)$ to denote the cumulative distribution function (CDF). 
$\erfc\left( \cdot \right)$ is used to denote the complementary error function given by $\erfc(x) = \frac{2}{\sqrt{\pi}} \int_{x}^{\infty}{e^{-u^2} du}$, $\erfcinv(\cdot)$ is the inverse of the complementary error function given by $\erfcinv(\erfc(x))=x$, and $\log (\cdot)$ is used to denote the logarithm with basis 2. 
We use $h(\cdot)$ to denote the entropy of a continuous RV and $I(\cdot;\cdot)$ to denote the mutual information between two RVs, as defined in \cite[Ch. 8.5]{cover-book}. 
We use $\styp(X)$ to denote the set of $\epsilon$-strongly typical sequences with respect to the probability mass function $p_X(x)$, as defined in \cite[Ch. 10.1]{cover-book}; when referring to a typical set we may omit the RVs from the notation, when these variables are clear from the context. Finally, $X \leftrightarrow Y \leftrightarrow Z$ is used to denote a Markov chain formed by the RVs $X,Y,Z$ as defined in \cite[Ch. 2.8]{cover-book}.

\vspace{-0.2cm}
\subsection{Molecular Timing Channel} \label{subsec:ALN}

We consider a molecular communication channel in which information is modulated on the time of release  of the information particles. This channel is illustrated in Fig. \ref{fig:diffuseMolComm}.  
The information particles themselves are assumed to be {\em identical and indistinguishable} at the receiver. Therefore, the receiver can only use the time of arrival to decode the intended message.
The information particles propagate from the transmitter to the receiver through some random propagation mechanism (e.g. diffusion). To develop our model, we make the following assumptions about the system:

\begin{enumerate}[label = {\bf A{\arabic*}})]
	\item \label{assmp:perfectTxRx}
	The transmitter perfectly controls the release time of each information particle, and the receiver perfectly measures the arrival times of the information particles. 
	Furthermore, the transmitter and the receiver are perfectly synchronized in time.		
	
	\item \label{assmp:Arrival}
	An information particle which arrives at the receiver is absorbed and hence is removed from the propagation medium.
	
	\item \label{assmp:indep}
	All information particles propagate independently of each other, and their trajectories are random according to an independent and identically distributed (i.i.d.)  random process. This is a fair assumption for many different propagation schemes in molecular communication such as diffusion in dilute solutions, i.e., when the number of particles released is much smaller than the number of molecules of the solutions.
\end{enumerate} 
\noindent Note that these assumptions have been adopted in all previous works \cite{ein2011,ein11ITW,ata13,nak12, cha12,eck12,pie13, li14, rose14} to make the models tractable.

Let $T_{x,k} \in \realSet^+, k=1,2,\dots,K$, denote the time of the $k^{\text{th}}$ transmission. At $T_{x,k}$, $M \in \intSet$ information particles are {\em simultaneously} released into the medium by the transmitter. 
The transmitted information is encoded in the sequence of times $\{T_{x,k}\}_{k=1}^K$, where $\{T_{x,k}\}_{k=1}^K$ are assumed to be independent of the random propagation time of {\em each} of the information particles.
Let $\vec{T}_{y,k}$ be an $M$-length vector consisting of the times of arrival of each of the information particles released at time $T_{x,k}$. Therefore, we have $\vec{T}_{y,k}[i] \geq T_{x,k}, i=1,2,\dots,M$. 
We further define $\vec{T}_{x,k}$ to be a vector consisting of $M$ repeated values of $T_{x,k}$. 
Thus, we obtain the following vector additive noise channel model:
\begin{align}
\label{eq:levyChan}
\vec{T}_{y,k} = \vec{T}_{x,k} + \vec{T}_{n,k}, 
\end{align}

\noindent where $\vec{T}_{n,k}[i], i=1,2,\dots,M$, is a random noise term representing the propagation time of the $i^{\text{th}}$ particle of the $k^{\text{th}}$ transmission.
Note that assumption \ref{assmp:indep} implies that all the elements of $\vec{T}_{n,k}$ are independent. 

%

One of the main challenges of the channel in \eqref{eq:levyChan} is that the particles may arrive out of order, which results in channel memory. To resolve this issue, we make two assumptions. First, we assume that at the beginning of each transmission there is a finite time interval called the {\em symbol interval} over which the transmitter can choose a time to release the information particles for that transmission. Second, we assume that information particles have a finite lifetime, i.e., they dissipate immediately after this finite interval, denoted by the {\em particle's lifetime}. 
By setting the channel use interval to be a concatenation of the symbol interval and the particle's lifetime, we ensure that order is preserved and obtain a memoryless channel.

Let $\tau_x < \infty$ be the symbol interval, and  $\tau_n < \infty$ be the particle's lifetime (i.e. each transmission interval is equal to $\tau_x+\tau_n$). Then our two assumptions can be formally stated as:
\begin{enumerate}[label = {\bf A{\arabic*}}), resume]
	\item	
	The release times obey:
	\begin{align*}
	(k-1)\cdot (\tau_x + \tau_n) \le T_{x,k} \leq (k-1)\cdot (\tau_x + \tau_n) + \tau_x.
	\end{align*}		
	
	\item \label{assmp:lifetime}
	The information particles dissipate and are never received if $\vec{T}_{n,k}[i] \geq \tau_n, i=1,2,\dots,M$. \label{asmp:limNoise}
	
\end{enumerate}

\begin{figure}
	\begin{center}
		\includegraphics[width=0.9\columnwidth,keepaspectratio]{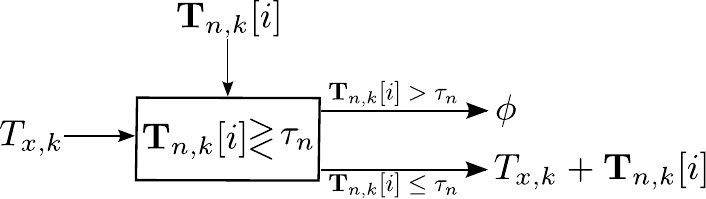}
	\end{center}
	\vspace{-0.3cm}
	\caption{\label{fig:chanModel} The MT channel in \eqref{eq:TCALNchan}. The channel input is $T_{x,k}$, while the channel output depends on the condition $T_{n,k} \gtrless \tau_n$.  }
\end{figure}

\noindent The first assumption can be justified by noting that the transmitter can choose its release interval, while the second assumption can be justified by designing the system such that information particles are degraded in the environment after a finite time (e.g. using chemical reactions) \cite{noe14,guo15}. 
The resulting channel, which we call the {\em molecular timing (MT) channel}, is given by:
\begin{align}
\label{eq:TCALNchan}
\vec{Y}_k[i] = \begin{cases} \vec{T}_{y,k}[i]= T_{x,k} + \vec{T}_{n,k}[i], & \vec{T}_{n,k}[i] \leq \tau_n \\ \phi, & \vec{T}_{n,k}[i] > \tau_n \end{cases},
\end{align}
where $T_{x,k}$ is the channel input, i.e., the $k^{\text{th}}$ release timing, $\vec{T}_{y,k}[i]$ is the arrival time of  the $i^{\text{th}}$ information particle at the receiver (if it arrives), and $\vec{Y}_k$ is an $M$-length vector of channel outputs at the $k^{\text{th}}$ channel use interval. The $i^{\text{th}}$ element of the MT channel \eqref{eq:TCALNchan} is depicted in Fig.~\ref{fig:chanModel}. Next, we formally define the capacity of the MT channel.

\vspace{-0.15cm}
\subsection{Capacity Formulation for the MT Channel} \label{subsec:ProbDef}

\begin{figure*}[!t]
	\begin{center}
		\includegraphics[width=0.55\textwidth,keepaspectratio]{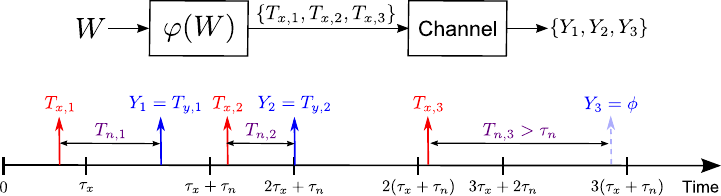}
	\end{center}
	\vspace{-0.3cm}
	\caption{\label{fig:Encoding} Illustration of the encoding procedure of Definition \ref{def:Encoding} for $K=3$ and $M=1$. Red pulses correspond to transmission times, while blue pulses correspond to arrival times at the receiver.}
	\vspace{-0.25cm}
\end{figure*}

Let $\mathcal{A}_k \triangleq [(k-1)\cdot (\tau_x+\tau_n), (k-1)\cdot (\tau_x+\tau_n) + \tau_x]$ and $\mathcal{B}_k \triangleq \left\{ [(k-1)\cdot (\tau_x+\tau_n), k\cdot (\tau_x+\tau_n)] \cup \phi \right\}$ for $k=1,2,\dots,K$.
We now define a code for the MT channel~\eqref{eq:TCALNchan} as follows:
\begin{definition}[Code] \label{def:Encoding}
	A $(K, R, \tau_x, \tau_n)$ code for the MT channel \eqref{eq:TCALNchan}, with code length $K$ and code rate $R$, consists of a message set $\mathcal{W} = \{1,2,\dots,2^{K(\tau_x+\tau_n)R} \}$, an encoder function $\varphi^{(K)}: \mathcal{W} \mapsto \mathcal{A}_1 \times \mathcal{A}_2 \times \dots \times \mathcal{A}_{K}$, and a decoder function  $\nu^{(K)}: \mathcal{B}_1^M \times \mathcal{B}_2^M \times \dots \times \mathcal{B}_{K}^M \mapsto \mathcal{W}$.
\end{definition}

\begin{rem}
	Observe that since we consider a timing channel, similarly to \cite{ana96}, the codebook size is a function of $\tau_x + \tau_n$, and $K(\tau_x + \tau_n)$ is the maximal time that it takes to transmit a message using a $(K, R, \tau_x, \tau_n)$ code. 
	Furthermore, note that the above encoder maps the message $W \in \mathcal{W}$ into $K$ time indices, $T_{x,k},k=1,2,\dots,K$, where $T_{x,k} \in \mathcal{A}_k$, while the decoder decodes the transmitted message using the $K \times M$ channel outputs $\{ \vec{Y}_k \}_{k=1}^K$ where $\vec{Y}_k \in \mathcal{B}_k^M$. We emphasize that this construction creates an {\em ordering} of the different arrivals, namely, each of the $M$ particles transmitted at the interval $\mathcal{A}_k$ either arrive {\em before} the $M$ particles transmitted at the interval $\mathcal{A}_{k+1}$ or will never arrive. Thus, we obtain $K$ identical and independent channels. Finally, we note that this construction was not used in \cite{ana96} since, when transmitting bits through queues, the channel itself forces an ordering.
\end{rem}

The encoding and transmission through the channel are illustrated in Fig. \ref{fig:Encoding} for the case of $K=3$ and $M=1$. The encoder produces three release times $\{T_{x,1}, T_{x,2}, T_{x,3}\}$ which obey $T_{x,k} \in \mathcal{A}_k, k=1,2,3$. In each time index a single particle is released to the channel which adds a random delay according to \eqref{eq:TCALNchan}. The channel outputs are denoted by $\{Y_1,Y_2,Y_3\}$. It can be observed that while $Y_1=T_{y,1}=T_{x,1}+T_{n,1}$ and $Y_2=T_{y,2}=T_{x,2}+T_{n,2}$, $Y_3 = \phi$ since $T_{n,3} > \tau_n$ and therefore the third particle does not arrive.

\begin{definition}[Probability of Error]
	The average probability of error of a $(K, R, \tau_x, \tau_n)$ code is defined as: 
	\begin{align*}
	P_e^{(K)} \triangleq \Pr \left\{ \nu(\mathcal{B}_1^M \times \mathcal{B}_2^M \times \dots \times \mathcal{B}_{K}^M) \neq W \right\},
	\end{align*}
	
	\noindent where the message $W$ is selected uniformly from the message set $\mathcal{W}$.
	
\end{definition}

\begin{definition}[Achievable Rate]
	A rate $R$ is called achievable if for any $\epsilon > 0$ and $\delta > 0$ there exists some blocklength $K_0(\epsilon, \delta)$ such that for every $K> K_0(\epsilon, \delta)$ there exits an $(K, R-\delta, \tau_x, \tau_n)$ code with $P_e^{(K)} < \epsilon$.
\end{definition}

\begin{definition}[Capacity]
	The capacity $\mathsf{C}$ is the supremum of all achievable rates.
\end{definition}

\begin{rem}
	Note that even though we consider a timing channel, we define the capacity in terms of bits per time unit \cite[Definition 2]{ana96}. This is in contrast to the works \cite{sri12, cha12,eck12,li14} which defined the capacity as the maximal number of bits which can be conveyed through the channel {\em per channel use}.
\end{rem}

Note that this definition of capacity $\mathsf{C}$ for the MT channels is fairly general and can be applied to different propagation mechanisms as long as Assumptions \ref{assmp:perfectTxRx}--\ref{assmp:lifetime} are not violated. Our objective in this paper is to characterize the capacity of the MT channel for the diffusion-based propagation.

\vspace{-0.2cm}
\subsection{Diffusion-Based MT Channel}
In diffusion-based propagation, the released information particles follow a random Brownian path from the transmitter to the receiver. In this case, to specify the random additive noise term $\vec{T}_{n,k}[i]$ in \eqref{eq:TCALNchan}, we define a L\'evy-distributed RV as follows: 
\begin{definition}[\levy Distribution] \label{def:levyRV}
	Let the RV $X$ be a L\'evy-distributed with location parameter $\mu$ and scale parameter $c$ \cite{nol15}. Then, its PDF is given by:
	\begin{align}
	\label{eqn:LevyPDF_0}
	f_X(x)=
	\begin{cases}
	\sqrt{\frac{c}{2 \pi (x-\mu)^3}}\exp \left( -\frac{c}{2(x-\mu)} \right), & x>\mu \\
	0, & x\leq \mu
	\end{cases},
	\end{align} 
	and its CDF is given by:
	\begin{align}
	\label{eqn:LevyCDF}
	F_X(x) = \begin{cases} \erfc\left(\sqrt{\frac{c}{2(x-\mu)}}\right), & x>\mu \\ 0, & x\leq\mu \end{cases}.		
	\end{align}
	The entropy of $X$, $h(x)$, is given by:
	\begin{align}
	\label{eq:entLevy}
	h(X) = \frac{\log(16c^2\pi e)+3\gamma\log(e)}{2},
	\end{align}
	where $\gamma \mspace{-3mu} \approx \mspace{-3mu} 0.5772$ is the Euler's constant \cite[Ch. 5.2]{nist10}. 
	Although this entropy is known, we did not find a rigorous proof in the literature, thus, the proof is provided in Appendix~\ref{app:ProofEntLevy}. Throughout the paper, we use the notation $X \sim \LevyDist(\mu,c)$ to indicate a \levy random variable with parameters $\mu$ and $c$. 
\end{definition}

Let $d$ denote the distance between the transmitter and the receiver, and $D$ denote the diffusion coefficient of the information particles in the propagation medium.   Following along the lines of the derivations in \cite[Sec. II]{sri12}, and using \cite[Sec. 2.6.A]{karatzas-shreve}, it can be shown that for the 1-dimensional pure diffusion, the propagation time of each of the information particles follows a \levy distribution, and therefore the noise in \eqref{eq:TCALNchan} is distributed as $\vec{T}_{n,k}[i] \sim \LevyDist(0,c)$ with $c = \frac{d^2}{2D}$. In this case, we call the channel in \eqref{eq:levyChan} the {\em additive \levy noise (ALN) channel}, and the MT channel in \eqref{eq:TCALNchan} the {\em DBMT channel}.

\begin{rem}
	In \cite{yilmaz20143dChannelCF} it is shown that for an infinite,  three-dimensional homogeneous medium without flow with a spherically absorbing receiver, the first arrival time follows a scaled \levy distribution. Therefore, the results presented in this paper can be extended to 3-D space by simply introducing a scalar multiple. 
\end{rem}


\section{The Capacity of the Single-Particle DBMT Channel} \label{sec:singlePart}

There are two main results in this section: Theorem \ref{thm:singlePartCap}, in which we obtain a general expression for the capacity of the single-particle DBMT channel; and Theorem \ref{th:UBLB}, in which we give closed-form upper and lower bounds on this capacity.

Since we study the capacity of the single-particle DBMT channel in \eqref{eq:TCALNchan} (i.e., when $M=1$), we use $Y_k$ instead of $\vec{Y}_k[i]$, $T_{y,k}$ instead of $\vec{T}_{y,k}[i]$, and $T_{n,k}$ instead of $\vec{T}_{n,k}[i]$. The channel \eqref{eq:TCALNchan} can now be written~as:
\begin{align}
\label{eq:TCALNchan_single}
	Y_k = \begin{cases} T_{y,k}= T_{x,k} + T_{n,k}, & T_{n,k} \leq \tau_n \\ \phi, & T_{n,k} > \tau_n \end{cases},
\end{align}

\noindent for $k=1,2,\dots,K$. Let $\mathcal{F}(\tau_x)$ denote the set of all PDFs $f_{T_x}(t_x)$ such that $F_{T_x}(t)=0$ for $t<0$ and $F_{T_x}(\tau_x)=1$.
The following theorem presents an expression for the capacity of the single-particle DBMT channel in~\eqref{eq:TCALNchan_single}. 
\begin{theorem} \label{thm:singlePartCap}
The capacity of the single-particle DBMT channel in (\ref{eq:TCALNchan_single}) is given by:
\begin{align}
\label{eq:TCALNchanCap}
 \mathsf{C}(\tau_n) \mspace{-3mu} =  \mspace{-3mu} \underset{\tau_x, \mathcal{F}(\tau_x)}{\max}  \frac{I(T_x;T_y|T_n<\tau_n)F_{T_n}(\tau_n)}{\tau_x + \tau_n}.
\end{align} 	
\end{theorem}

\begin{IEEEproof}
	In Appendix \ref{annex:BasicCapacityProof} we show that the capacity of the channel \eqref{eq:TCALNchan_single}, in bits per second, is given by:
	\begin{align}
		\label{eq:capacityDef}
		\mathsf{C}(\tau_n) = \underset{\tau_x, \mathcal{F}(\tau_x)}{\max} \frac{I(T_x; Y)}{\tau_x + \tau_n}.
	\end{align}
	
 Note that the channel \eqref{eq:TCALNchan_single} implies that $Y_k$ does not have a density, and therefore a straight-forward evaluation of $I(T_x, Y)$ via a simple integration cannot be applied. 
	To evaluate \eqref{eq:capacityDef}, we first note that the channel model in (\ref{eq:TCALNchan_single}) can be represented as two separate channel models, where at each channel use only one of the channels is selected at random for transmission. This is illustrated in Fig. \ref{fig:chanModel}. Let $\Theta$ be a Bernoulli random variable that indicates which channel is selected at random:
	\begin{align}
	\Theta=
	\begin{cases}
	1, & T_n \leq \tau_n \\
	0, & T_n > \tau_n 
	\end{cases}. \label{eq:thetaDef}
	\end{align}
	
	\noindent Hence, $\Theta$ has a probability of success $p=F_{T_n}(\tau_n)$. 
	Since for each case the received symbol sets are disjoint, we have the Markov chain $T_x \leftrightarrow Y \leftrightarrow \Theta$. We next write:
	\begin{align}
		I(T_x;Y) &= I(T_x;Y,\Theta) \label{eqn:baseMI_markov} \\
				&=I(T_x;\Theta)+I(T_x;Y|\Theta) \nonumber \\
				&=I(T_x;Y|\Theta) \label{eqn:baseMI_MITxThetaZero} \\
				&=\Pr \{ \Theta=1 \} \cdot I(T_x;Y|\Theta=1) \nonumber \\
				& \qquad + \Pr \{\Theta=0 \} \cdot I(T_x;Y|\Theta=0) \nonumber\\
				&=\Pr \{\Theta=1\} \cdot I(T_x;T_y|\Theta=1), \label{eqn:baseMI_removeNull}
		\end{align}
		
		\noindent where \eqref{eqn:baseMI_markov} follows from the Markov chain $T_x \leftrightarrow Y \leftrightarrow \Theta$; \eqref{eqn:baseMI_MITxThetaZero} follows from the fact that the channel input is independent of the selected channel, which is a function only of the additive noise; and \eqref{eqn:baseMI_removeNull} follows from the fact that when $\Theta =0$, no information goes through the channel and therefore $I(T_x;\phi|\Theta \mspace{-3mu} = \mspace{-3mu} 0) \mspace{-3mu} = \mspace{-3mu} 0$.
		Finally, we note that \eqref{eq:thetaDef} implies $I(T_x;T_y|\Theta \mspace{-3mu} = \mspace{-3mu} 1) \mspace{-3mu}  = \mspace{-3mu} I(T_x;T_y|T_n \mspace{-3mu} \leq \mspace{-3mu} \tau_n)$, and $\Pr \{\Theta=1\} \mspace{-3mu} =$ $\Pr \{T_n \leq \tau_n\} = F_{T_n}(\tau_n)$; thus, we obtain~\eqref{eq:TCALNchanCap}.
\end{IEEEproof}


Obtaining an exact expression for \eqref{eq:TCALNchanCap} is highly complicated as the maximizing input distribution $f_{T_x}(t_x) \in \mathcal{F}(\tau_x)$ is not known. Therefore, we turn to upper and lower bounds. 
We first note that the conditional mutual information in (\ref{eq:TCALNchanCap}) can be written~as:
\begin{align}
	I(T_x;T_y| T_n  \leq \tau_n) \mspace{-3mu} &= \mspace{-3mu} h(T_y| T_n \mspace{-3mu} \leq \mspace{-3mu} \tau_n) \mspace{-3mu} - \mspace{-3mu} h(T_y|T_x, T_n \leq \tau_n) \nonumber \\
	& = \mspace{-3mu} h(T_y| T_n \leq  \tau_n) \mspace{-3mu} - \mspace{-3mu} h(T_n| T_n  \leq  \tau_n), \label{eq:Ty_is_a_sum}
\end{align}

\noindent where \eqref{eq:Ty_is_a_sum} follows from the fact that $T_y = T_x + T_n$ for $T_n \le \tau_n$.
In the following we explicitly evaluate $h(T_n| T_n \leq \tau_n)$ and bound $h(T_y| T_n  \leq \tau_n)$.

\subsection{Characterizing $h(T_n| T_n \leq \tau_n)$}

To characterize the conditional entropy $h(T_n| T_n \leq \tau_n)$ we first define the {\em partial entropy} of a continuous RV $X$, which captures the entropy of the continuous RV in the range $(-\infty, \tau]$:
	\begin{definition}[Partial Entropy]
		The partial entropy of a random variable $X$ with PDF $f(x)$ and parameter $\tau$ is defined~by:
		\begin{align}
		\label{eq:funcN}
			\eta(X,\tau) = - \int_{-\infty}^\tau f(x) \log(f(x)) dx.
		\end{align}
	\end{definition}
	
	Let $X$ be a continuous RV with PDF $f_X(x)$ and CDF $F_X(x)$, and let $\tau$ be a real constant. The following theorem uses the above definition to characterize $h(X | X < \tau)$:
		\begin{theorem}
			\label{th:entrCondLev}	
			The conditional entropy $h(X|X \le \tau)$ of a continuous RV $X$ is given by:
			\begin{align}
			\label{eq:condEntLev}
			h(X|X \le \tau) = \frac{\eta(X,\tau)}{F_X(\tau)} + \log(F_X(\tau)),
			\end{align}
			where $\eta(X,\tau)$ is the partial entropy.
		\end{theorem}
		
		\begin{IEEEproof}
			We first note that the RV $\tilde{X}$, defined as $X$ given $X \le \tau$, has PDF $f_{\tilde{X}}(\tilde{x}) = \frac{f_X(x)}{F_X(\tau)}$. 			
			Next, we write the entropy of~$\tilde{X}$: 
			\begin{align}
				h(\tilde{X}) & = h(X|X<\tau) \nonumber \\
				&= -\int_{-\infty}^\tau \frac{f_X(x)}{F_X(\tau)} \log\left( \frac{f_X(x)}{F_X(\tau)}\right) dx \label{eqn:timeConstEntropy_1}\\
				&= -\frac{1}{F_X(\tau)} \int_{-\infty}^\tau f_X(x) \log(f_X(x))dx  \nonumber\\
				& \qquad +\frac{1}{F_X(\tau)} \int_{-\infty}^\tau f_X(x) \log (F_X(\tau)) dx \nonumber \\
				&= -\frac{1}{F_X(\tau)} \int_{-\infty}^\tau f_X(x) \log(f_X(x)) dx +\log(F_X(\tau)) \label{eqn:timeConstEntropy_2} \\
				&=\frac{\eta(X,\tau)}{F_X(\tau)} + \log(F_X(\tau)), \label{eqn:timeConstEntropy_3}
			\end{align}
			
			\noindent where \eqref{eqn:timeConstEntropy_1} follows from the definition of entropy; \eqref{eqn:timeConstEntropy_2} follows by noting that $\int_{-\infty}^{\tau}{f_X(x)dx} = F_X(\tau)$; and \eqref{eqn:timeConstEntropy_3} follows from the definition of $\eta(X,\tau)$.			
		\end{IEEEproof}
		
		As can be seen from Theorem \ref{th:entrCondLev}, to find an expression for the conditional entropy $h(X|X \le \tau)$, for a L\'evy-distributed RV $X$, one needs to find the partial entropy of $X$ (with offset parameter $\mu = 0$). This partial entropy is presented in the following lemma:
	\begin{lemma}
		\label{lem:partEntLevy}
		If $X \sim \LevyDist(0,c)$, then
		\begin{align}
		\label{eq:funcNlevy}
		\eta(X,\tau) &=\tfrac{1}{2}\log(\tfrac{2 \pi}{c})F_X(\tau)+\tfrac{3}{2}\bigg[(F_X(\tau)-1)\log(\tau)- \nonumber \\
		&  ~~~~4\sqrt{\tfrac{c}{2\pi\tau}}g(c,\tau)\log(e)+\log(c/2)+\gamma\log(e)+2\bigg] \nonumber\\
		&~~~~+ \log(e)\bigg[\tfrac{1}{2}F_X(\tau)+\tau f_X(\tau)\bigg],
		\end{align}
		
		\noindent where $f_X(x)$ is given in \eqref{eqn:LevyPDF_0}, $F_X(x)$ is given in \eqref{eqn:LevyCDF}, and $g(c,\tau)$ is a generalized hypergeometric function \cite[Ch. 16]{nist10} given by
		\begin{align}
		g(c,\tau) \triangleq~_2F_2(\tfrac{1}{2},\tfrac{1}{2},\tfrac{3}{2},\tfrac{3}{2};\tfrac{-c}{2\tau}).
		\end{align}
	\end{lemma}
	
	\begin{IEEEproof}
		The proof is provided in Appendix \ref{app:ProofPartEntLevy}. 
	\end{IEEEproof}
	
	To find $h(T_n|T_n \le \tau_n)$ we plug \eqref{eqn:LevyPDF_0} and \eqref{eqn:LevyCDF} into \eqref{eq:funcNlevy}, and then plug the resulting expression into \eqref{eqn:timeConstEntropy_3}.
	
	\subsection{Bounds on the Capacity}

Since the maximizing input distribution in \eqref{eq:TCALNchanCap} is not known, it is difficult to obtain an exact expression for the maximal value of $h(T_y | T_n \le \tau_n)$. Therefore, we turn to lower and upper bounds on $h(T_y | T_n \le \tau_n)$, which results in lower and upper bounds on $\mathsf{C}(\tau_n)$. 
For the lower bound we note that $h(T_y | T_n \le \tau_n) = h(T_x + T_n | T_n \le \tau_n)$ and use the entropy power inequality (EPI) \cite[pg. 22]{kim-elgamal-book} to obtain a bound in terms of $h(T_x)$ and $h(T_n | T_n \le \tau_n)$.\footnote{The work \cite{khormuji11} was the first to use the EPI in deriving a lower bound on the capacity of MT channels.} For the upper bound we again use the relationship $T_y = T_x + T_n$ to bound $h(T_y | T_n \le \tau_n)$ by the logarithm of the support of $T_y$.
Define $m(\tau_x,\tau_n,T_n)$ as:
\begin{align}
		m(\tau_x,\tau_n,T_n) = 0.5 \log \big( \tau_x^2+2^{2h(T_n|T_n\leq\tau_n)}\big),
		\label{eqn:mFuncDef}
	\end{align} 
	
\noindent and recall that $h(T_n|T_n\leq\tau_n)$ is characterized in Theorem \ref{th:entrCondLev}. 
The following theorem presents the lower and upper bounds on $\mathsf{C}(\tau_n)$:
\begin{theorem}
	\label{th:UBLB}
	The capacity of the single-particle DBMT channel is bounded by $\mathsf{C}^{\text{lb}}(\tau_n) \le \mathsf{C}(\tau_n) \le \mathsf{C}^{\text{ub}}(\tau_n)$, where $\mathsf{C}^{\text{lb}}(\tau_n)$ and $\mathsf{C}^{\text{ub}}(\tau_n)$ are given by:
	\begin{align}
	\mathsf{C}^{\text{lb}}(\tau_n) & \mspace{-1mu} \triangleq \mspace{-1mu} \underset{\tau_x}{\max} \frac{\left(m(\tau_x,\tau_n,T_n) \mspace{-3mu} - \mspace{-3mu} h(T_n|T_n\leq\tau_n)\right)F_{T_n}(\tau_n)}{\tau_x + \tau_n} \label{eq:capacityLB_single} \\
	\mathsf{C}^{\text{ub}}(\tau_n) & \mspace{-1mu} \triangleq \mspace{-1mu} \underset{\tau_x}{\max} \frac{\left(\log(\tau_x + \tau_n) \mspace{-3mu} - \mspace{-3mu} h(T_n|T_n\leq\tau_n)\right)F_{T_n}(\tau_n)}{\tau_x+\tau_n}. \label{eq:capacityUB_single}
	\end{align}
\end{theorem}

\begin{IEEEproof}
	For the lower bound $\mathsf{C}^{\text{lb}}(\tau_n)$ we write:
	\begin{align}
		h(T_y| T_n \leq \tau_n) &= h(T_x+T_n| T_n \leq \tau_n) \nonumber \\
								&\geq 0.5 \log \bigg( 2^{2h(T_x|T_n \leq \tau_n)}+2^{2h(T_n|T_n \leq \tau_n)}\bigg) \label{eq:powerInq} \\
											&= 0.5 \log \bigg( 2^{2h(T_x)}+2^{2h(T_n|T_n \leq \tau_n)}\bigg), \label{eq:powerInq_2}
	\end{align}
	where \eqref{eq:powerInq} follows from the EPI, and \eqref{eq:powerInq_2} follows by noting that $T_x$ and $T_n$ are independent given $T_n \leq \tau_n$.
	Furthermore, as this bound holds for every $f_{T_x}(t_x)$, we use the entropy maximizing distribution for $T_x$, the uniform distribution, with entropy $\log(\tau_x)$ to obtain $m(\tau_x,\tau_n,T_n)$.
	
	For the upper bound $\mathsf{C}^{\text{ub}}(\tau_n)$ we write:
		\begin{align}
		h(T_y| T_n \leq \tau_n) &\leq  \log(\tau_x + \tau_n), \label{eqn:capacityUB_2}
		\end{align}
		
		\noindent where  \eqref{eqn:capacityUB_2} follows since given the event $T_n \leq \tau_n$, $0<T_y\leq \tau_x + \tau_n$, and the uniform distribution maximizes entropy over a finite interval.
\end{IEEEproof}

Let $\varepsilon(\tau_n) \triangleq 2^{1 + h(T_n|T_n\leq\tau_n)}$.
The following corollary provides an explicit solution to the maximization problem in \eqref{eq:capacityUB_single}:
\begin{cor} \label{cor:explicitUB_value}
	An explicit solution for the maximization problem defined in \eqref{eq:capacityUB_single} is given by:
	\begin{align*}
		\mathsf{C}^{\text{ub}}(\tau_n) \mspace{-3mu} = \mspace{-3mu}
			\begin{cases} \frac{F_{T_n}(\tau_n)}{\varepsilon(\tau_n)}, & \mspace{-3mu} \varepsilon(\tau_n) \mspace{-2mu} > \mspace{-2mu} \tau_n \\
			(\log \left( \tau_n \right) \mspace{-3mu} - \mspace{-3mu} h(T_n|T_n\leq\tau_n)) \frac{F_{T_n}(\tau_n)}{\tau_n}, & \mspace{-3mu} \varepsilon(\tau_n) \mspace{-2mu} \le \mspace{-2mu} \tau_n \end{cases},
	\end{align*}
	
	\noindent where the maximizing $\tau_x$ is given by $\tau_x^{\ast} = \max \{0, \varepsilon(\tau_n) - \tau_n\}$. Furthermore, the $\tau_x$ which maximizes \eqref{eq:capacityLB_single} is a solution of the following equation in $\tau_x$:
	\begin{align*}
		& h(T_n|T_n \mspace{-3mu} \leq \mspace{-3mu} \tau_n) \left(4^{h(T_n|T_n \mspace{-3mu} \leq \mspace{-3mu} \tau_n)} + \tau_x^2 \right) + \tau_x (\tau_x + \tau_n) \nonumber \\
		&\quad - \frac{1}{2} \left(4^{h(T_n|T_n \mspace{-3mu} \leq \mspace{-3mu} \tau_n)} + \tau_x^2 \right) \log \left(4^{h(T_n|T_n \mspace{-3mu} \leq \mspace{-3mu} \tau_n)} + \tau_x^2 \right) = 0.
	\end{align*}
\end{cor}

\begin{IEEEproof}
	The proof is provided in Appendix \ref{annex:cor_explicitUB_value_proof}.
\end{IEEEproof}

\begin{rem} \label{rem:argumentsDiverge}
	For $\tau_n \to \infty$, the maximizing $\tau_x$'s for the bounds in \eqref{eq:capacityLB_single} and \eqref{eq:capacityUB_single} diverge. To see this, we first note that $\lim_{\tau_n \to \infty} h(T_n|T_n \mspace{-3mu} \leq \mspace{-3mu} \tau_n) \mspace{-3mu} = \mspace{-3mu} h(T_n) \mspace{-3mu} < \mspace{-3mu} \infty$, given in \eqref{eq:entLevy}. Hence, Corollary \ref{cor:explicitUB_value} implies that when $\tau_n \to \infty$ then \eqref{eq:capacityUB_single} is maximized by $\tau_x = 0$. On the other hand, the maximizing $\tau_x$ for \eqref{eq:capacityLB_single} is a solution of the following equation:
	\begin{align*}
		h(T_n) + \frac{\tau_x (\tau_x + \tau_n)}{4^{h(T_n)} + \tau_x^2} = \log (4 + \tau_x^2).
	\end{align*}
	
\noindent The two possible solutions for this equation are $\tau_x \to 0$ and $\tau_x \to \infty$. Since when $\tau_x = 0$ we have $\mathsf{C}^{\text{lb}}(\tau_n) = 0$, regardless of the value of $\tau_n$, 
we conclude that the maximizing $\tau_x$ tends to infinity.
\end{rem}

\begin{rem}
For $\tau_n \to \infty$ the capacity $\mathsf{C}(\tau_n) \to 0$. Intuitively, $\tau_n$ can be viewed as a guard interval that insures ordered arrivals. Clearly, if such a guard interval is infinite, the capacity is zero. This can be formally justified by writing the upper bound \eqref{eq:capacityUB_single}, for $\tau_n \to \infty$, as:
\begin{align*}
	& \lim_{\tau_n \to \infty} \underset{\tau_x}{\max} \frac{\left(\log(\tau_x + \tau_n) \mspace{-3mu} - \mspace{-3mu} h(T_n|T_n\leq\tau_n)\right)F_{T_n}(\tau_n)}{\tau_x+\tau_n}  \nonumber \\
	& \qquad = \lim_{\tau_n \to \infty} \underset{\tau_x}{\max} \frac{\left(\log(\tau_x + \tau_n) \mspace{-3mu} - \mspace{-3mu} h(T_n)\right)}{\tau_x+\tau_n} \nonumber \\
		& \qquad = 0,
\end{align*}

\noindent where the last equality follows from the fact that $h(T_n)$ is finite and $\tau_x > 0$.
\end{rem}

\begin{rem} \label{rem:argumentsConverge}
	For a fixed $\tau_n$ and $\tau_x \to \infty$, the arguments of the maximization problems in \eqref{eq:capacityLB_single} and \eqref{eq:capacityUB_single} converge, namely:
	\begin{align*}
		\lim_{\tau_x \to \infty} \frac{\log(\tau_x + \tau_n) \mspace{-3mu} - \mspace{-3mu} h(T_n|T_n\leq\tau_n)}{m(\tau_x,\tau_n,T_n) \mspace{-3mu} - \mspace{-3mu} h(T_n|T_n\leq\tau_n)} = 1.
	\end{align*}
	
	\noindent This follows from the fact that for $\tau_x \mspace{-3mu} \gg \mspace{-3mu} 4^{h(T_n|T_n \mspace{-3mu} \leq \mspace{-3mu} \tau_n)}$ we have $m(\tau_x,\tau_n,T_n) \approx \log(\tau_x)$.
\end{rem}

	%
%


\section{The Capacity of the DBMT Channel with Diversity}
\label{sec:multiPart}

	\begin{figure*}[!t]
	\normalsize
	\setcounter{MYtempeqncnt}{\value{equation}}
	\setcounter{equation}{27}
	
	\begin{align}
	\label{eq:mTCALNchanCap}
	\mathsf{C}_M(\tau_n) =  \underset{\tau_x, \mathcal{F}(\tau_x)}{\max} \left\{ \frac{1}{\tau_x + \tau_n} \displaystyle\sum_{\substack{\mathcal{J} = \{0,1,\dots,\tilde{J}\}: \\ \tilde{J} \in \{1,2,\dots,M\}}} I\left(\vec{T}_x[\mathcal{J}];\vec{T}_y[\mathcal{J}] \big| \vec{T}_n[\mathcal{J}] \le \tau_n \right) \cdot v(p, M, |\mathcal{J}|) \right\}.
	\end{align} 
	
	\hrulefill
	
	\setcounter{equation}{\value{MYtempeqncnt}}
	
\end{figure*}

The focus of Section \ref{sec:singlePart} is on the single-particle DBMT channel. i.e., $M=1$. In this section we address the question: {\em Can one improve performance by simultaneously releasing multiple particles, namely, using $M>1$ particles?} 
In \cite{murinGlobeCom16} and \cite[Sec. IV.C]{sri12} it is shown that by releasing multiple particles one can reduce the probability of error; yet, it is not clear if and how the capacity scales with the number of particles that are {\em simultaneously} released in each transmission interval $\mathcal{A}_k$ (see Section \ref{subsec:ProbDef} for the detailed definitions).\footnote{Note that simultaneously releasing multiple particles is analogous to receiver diversity as each particle follows an independent path from the transmitter to the receiver.} In the current and subsequent sections we investigate this problem.


The following is a road-map to our results for the DBMT channel with diversity. 
In Theorem \ref{thm:multPartCap}, we present a capacity expression for the case where the receiver {\em accurately} measures the arrival time of {\em all the particles}. Since analytic evaluation of this expression seems intractable, in Theorem \ref{th:mUB}, we present a capacity upper bound that scales linearly with the number of released particles $M$. 
Subsequently, in the Section \ref{sec:FA_AVG}, we consider two specific receivers; demonstrate that asymptotically (as $M\rightarrow\infty$) these receivers can be represented by additive noise channels where the noise terms are Gumbel or Gaussian distributed; and present upper and lower bounds on the capacity of the resulting two channel models. 
In particular, in Theorem \ref{th:FA_UBLB}, we provide bounds on the capacity of the system whose receiver measures the first arrival (FA) time of the particles, and show that the lower bound on this capacity, in {\em bits per channel use}, scales as $\log(\log(M))$, see Corollary \ref{cor:rateFA}. 
However, as we show Section \ref{sec:NumEvaluation}, optimizing over $\tau_x$ and $\tau_n$ may result in a poly-logarithmic scaling of the lower bound in {\em bits per second}. 
In Theorem \ref{th:AVG_UBLB}, we provide bounds on the capacity of the system whose receiver measures the average arrival time. The resulting lower bound, in {\em bits per channel use}, scales as $\log(M)$, see Corollary \ref{cor:rateAvg}. Again, as shown in Section \ref{sec:NumEvaluation}, optimizing over $\tau_x$ and $\tau_n$ may result in a poly-logarithmic scaling of the lower bound in {\em bits per second}. Clearly, these lower bounds are also lower bounds on the capacity of the system where the receiver accurately measures the arrival time of all the particles, i.e., it is a lower bound on the capacity of the DBMT channel without any constraints on the receiver.

\subsection{Capacity Expression for the General DBMT Channel with Diversity}

We begin our analysis with defining the set $\mathcal{J}_k \triangleq \{ j: \vec{T}_{n,k}[j] \le \tau_n\}, k=1,2,\dots,K$, which is the set of the indices of all particles which arrived within the interval $[(k-1)\cdot (\tau_x+\tau_n), k\cdot (\tau_x+\tau_n)]$. Clearly, $|\mathcal{J}_k| \le M$. Note that for every $0 \le l \le M, l \notin \mathcal{J}_k$, the output of the channel \eqref{eq:TCALNchan} is $\phi$, and therefore this particle does not convey information over the channel. More precisely, let $\vec{Y}_{k,\mathcal{J}_k}$ denote the vector $\vec{Y}_k[j], j \in \mathcal{J}_k$, and $\vec{Y}_{k,\mathcal{J}_k^c}$ denote the vector $\vec{Y}_k[l], l \notin \mathcal{J}_k$. We write:
\begin{align*}
	I(T_{x,k}; \vec{Y}_k) & = I(T_{x,k}; \vec{Y}_{k,\mathcal{J}_k}, \vec{Y}_{k,\mathcal{J}^c_k}) \\
	& = I(T_{x,k}; \vec{Y}_{k,\mathcal{J}_k}, [\phi, \phi, \dots, \phi]) \\
	& = I(T_{x,k}; \vec{Y}_{k,\mathcal{J}_k}).
\end{align*}

\noindent Since all the particles are statistically indistinct, the term $I(T_{x,k}; \vec{Y}_{k,\mathcal{J}_k})$ depends on $|\mathcal{J}_k|$ and not on the specific indices of the set $\mathcal{J}_k$. 
In fact, one can re-label the transmitted particles such that the first $|\mathcal{J}_k|$ are the particles that arrive within the interval $[(k-1)\cdot (\tau_x+\tau_n), k\cdot (\tau_x+\tau_n)]$. 
Therefore, in the following we slightly abuse the notation and let $\mathcal{J}_k = \{ 1,2,\dots,|\mathcal{J}_k| \}$. We define $\vec{T}_{y,k}[\mathcal{J}_k] \triangleq [\vec{T}_{y,k}[1], \vec{T}_{y,k}[2], \dots, \vec{T}_{y,k}[|\mathcal{J}_k|]]$, while $\vec{T}_{n,k}[\mathcal{J}_k]$ is defined in a similar manner. 
Finally, we define $\vec{T}_{x,k}[\mathcal{J}_k]$ to be a vector of length $|\mathcal{J}_k|$ with all its elements equal to the repeated values $T_{x,k}$. 
With this notation we now define a channel equivalent to \eqref{eq:TCALNchan}:
\begin{align}
\label{eq:mTCALNchan}
\vec{Y}_k=
\begin{cases}
\vec{T}_{y,k}[\mathcal{J}_k] \mspace{-3mu} = \mspace{-3mu} \vec{T}_{x,k}[\mathcal{J}_k] \mspace{-3mu} + \mspace{-3mu} \vec{T}_{n,k}[\mathcal{J}_k], & |\mathcal{J}_k| > 0 \\
\phi, &  |\mathcal{J}_k| = 0
\end{cases}.
\end{align}


Let $\mathsf{C}_M(\tau_n)$ denote the capacity of the DBMT channel with diversity in \eqref{eq:TCALNchan}, and therefore also the capacity of the channel \eqref{eq:mTCALNchan}. In addition, let $p \triangleq F_{T_n}(\tau_n)$, and define the function $v(p, M, i) \triangleq {M \choose i} p^{i}(1-p)^{M-i}, i=1,2,\dots,M$.
The following theorem characterizes $\mathsf{C}_M(\tau_n)$:

\begin{theorem} \label{thm:multPartCap}
	$\mathsf{C}_M(\tau_n)$ is given by \eqref{eq:mTCALNchanCap} at the top of the page, where the condition $\vec{T}_n[\mathcal{J}] \le \tau_n$ reads $\vec{T}_n[j] \le \tau_n, \forall j \in \mathcal{J}, \vec{T}_n[l] > \tau_n, \forall l \notin \mathcal{J}$.
	
	
\end{theorem}

\setcounter{equation}{28}

\begin{IEEEproof}
	We follow steps similar to those used in the proof of Theorem \ref{thm:singlePartCap}. Extending the proof detailed in Appendix \ref{annex:BasicCapacityProof}, 
	one can show that the capacity of the channel \eqref{eq:TCALNchan}, and therefore also the channel \eqref{eq:mTCALNchan}, in bits per second, is given by:
	\begin{align*}
		\mathsf{C}(\tau_n) = \underset{\tau_x, \mathcal{F}(\tau_x)}{\max} \frac{I(T_x; \vec{Y})}{\tau_x + \tau_n}.
	\end{align*}
	
	Next, we note that since the propagation of the different particles is independent, see assumption \ref{assmp:indep}, $|\mathcal{J}|$ follows a binomial distribution, i.e., $|\mathcal{J}| \sim \mathscr{B} (M,F_{T_n}(\tau_n))$.
		Furthermore, as $|\mathcal{J}|$ is a function of only the received symbol set $\vec{Y}$, we have the Markov chain $\vec{T}_x \leftrightarrow \vec{Y} \leftrightarrow |\mathcal{J}|$. Thus, we write:
	\begin{align}
		I(T_x;\vec{Y}) & = \mspace{-3mu} I(\vec{T}_x;\vec{Y}) \label{eqn:baseMI_mult} \\
		& = \mspace{-3mu} I(\vec{T}_x;\vec{Y},|\mathcal{J}|) \label{eqn:baseMI_Markov_mult} \\
		& = \mspace{-3mu} I(\vec{T}_x;\vec{Y} \big| |\mathcal{J}|) \label{eqn:baseMI_MITxThetaZero_mult} \\
		& = \mspace{-3mu} \sum_{j=0}^{M} \Pr \{|\mathcal{J}| \mspace{-3mu} = \mspace{-3mu} j \} \mspace{-3mu} \cdot \mspace{-3mu} I(\vec{T}_x[\mathcal{J}];\vec{T}_y[\mathcal{J}] \big| |\mathcal{J}| \mspace{-3mu} = \mspace{-3mu} j) \nonumber \\
		& = \mspace{-3mu} \sum_{j=1}^{M} \Pr \{|\mathcal{J}| \mspace{-3mu} = \mspace{-3mu} j \} \mspace{-3mu} \cdot \mspace{-3mu} I(\vec{T}_x[\mathcal{J}];\vec{T}_y[\mathcal{J}] \big| |\mathcal{J}| \mspace{-3mu} = \mspace{-3mu} j), \label{eqn:baseMI_fnl_mult}
	\end{align}
	
	\noindent where \eqref{eqn:baseMI_mult} follows from the fact that $\vec{T}_x$ is simply a vector which contains $T_x$ multiple times; \eqref{eqn:baseMI_Markov_mult} follows from the Markov chain $\vec{T}_x \leftrightarrow \vec{Y} \leftrightarrow |\mathcal{J}|$; \eqref{eqn:baseMI_MITxThetaZero_mult} follows from the fact that $\vec{T}_x$ is independent of $|\mathcal{J}|$; and, \eqref{eqn:baseMI_fnl_mult} follows by noting that when $|\mathcal{J}|=0$, no information is conveyed through the channel.
	
	Finally, we note that the condition $\vec{T}_n[\mathcal{J}] \le \tau_n, |\mathcal{J}|=j$ is equivalent to the condition $|\mathcal{J}| = j$, and since $|\mathcal{J}| \sim \mathscr{B} (M,F_{T_n}(\tau_n))$ then $\Pr \{|\mathcal{J}| = j \} = v(p, M, j)$.
\end{IEEEproof}

\subsection{An Upper Bound}

Similarly to the single-particle case, obtaining an exact expression for $\mathsf{C}_M(\tau_n)$ is highly complicated, thus, we turn to upper and lower bounds. 
The next theorem provides an upper bound on the capacity in \eqref{eq:mTCALNchanCap}.
\begin{theorem}
	\label{th:mUB}
	The capacity of the DBMT channel with diversity is upper bounded by $\mathsf{C}_M(\tau_n) \le \mathsf{C}_M^{\text{ub}}(\tau_n)$, where $\mathsf{C}_M^{\text{ub}}(\tau_n)$ is given by:
	\begin{align}
	\mathsf{C}_M^{\text{ub}}(\tau_n) & \mspace{-3mu} \triangleq \mspace{-3mu} \max_{\tau_x} \frac{\left(\log(\tau_x + \tau_n) \mspace{-3mu} - \mspace{-3mu} h(T_n|T_n\leq\tau_n)\right) \mspace{-3mu} \cdot \mspace{-3mu} M \mspace{-3mu} \cdot \mspace{-3mu} F_N(\tau_n)}{\tau_x + \tau_n}, \label{eq:capacityUB_mul}
	\end{align} 
	
	\noindent and $h(T_n|T_n\leq\tau_n)$ is given in Theorem \ref{th:entrCondLev}.
\end{theorem}	


\begin{IEEEproof}
	First, we note that the conditional mutual information in \eqref{eq:mTCALNchanCap} can be written as:
\begin{align}
	& I\left(\vec{T}_x[\mathcal{J}];\vec{T}_y[\mathcal{J}] \big| \vec{T}_n[\mathcal{J}] \le \tau_n \right) \nonumber \\
	& \qquad \qquad = h\left(\vec{T}_y[\mathcal{J}] \big| \vec{T}_n[\mathcal{J}] \le \tau_n \right) \nonumber \\
	& \qquad  \qquad \qquad - h\left(\vec{T}_y[\mathcal{J}] \big| \vec{T}_x[\mathcal{J}], \vec{T}_n[\mathcal{J}] \le \tau_n \right) \\
	& \qquad  \qquad = h\left(\vec{T}_y[\mathcal{J}] \big| \vec{T}_n[\mathcal{J}] \le \tau_n \right) \nonumber \\
	& \qquad \qquad  \qquad- h\left(\vec{T}_n[\mathcal{J}] \big| \vec{T}_n[\mathcal{J}] \le \tau_n \right).
\end{align} 

\noindent Next, we explicitly evaluate $h\left(\vec{T}_n[\mathcal{J}] \big| \vec{T}_n[\mathcal{J}] \le \tau_n \right)$ and bound $h\left(\vec{T}_y[\mathcal{J}] \big| \vec{T}_n[\mathcal{J}] \le \tau_n \right)$. From assumption \ref{assmp:indep} we have:
\begin{align}
	h\left(\vec{T}_n[\mathcal{J}] \big| \vec{T}_n[\mathcal{J}] \le \tau_n \right) & = \sum_{j=1}^{|\mathcal{J}|} h\left(\vec{T}_n[j] \big| \vec{T}_n[j] \le \tau_n \right) \nonumber \\
	& = |\mathcal{J}| \cdot h\left(T_n | T_n \le \tau_n \right).
\end{align}
	
	\noindent Next, we bound $h\left(\vec{T}_y[\mathcal{J}] \big| \vec{T}_n[\mathcal{J}] \le \tau_n \right)$ as
	\begin{align}
	h(\vec{T}_y[\mathcal{J}]|\vec{T}_n[\mathcal{J}]<\tau_n) & \leq h(\vec{T}_x[\mathcal{J}] + \vec{T}_n[\mathcal{J}]) \label{eq:mult_upper_prf_1} \\
	& \leq \sum_{j=1}^{|\mathcal{J}|} h(T_x + T_n) \label{eq:mult_upper_prf_2} \\
	&\leq |\mathcal{J}| \log(\tau_x + \tau_n), \label{eq:mult_upper_prf_3}
	\end{align}
	\noindent where \eqref{eq:mult_upper_prf_1} and \eqref{eq:mult_upper_prf_2} follow from the fact that conditioning reduces entropy, and \eqref{eq:mult_upper_prf_3} is due to the fact that the uniform distribution maximizes entropy over a finite interval. 
	Therefore \eqref{eq:mTCALNchanCap} can be upper bounded by:
	\begin{align} 
	\mspace{-8mu} \sum_{j=1}^{M} (\log(\tau_x + \tau_n) \mspace{-3mu} - \mspace{-3mu} h(T_n|T_n\leq\tau_n)) \mspace{-3mu} \cdot \mspace{-3mu} j \mspace{-3mu} \cdot \mspace{-3mu} v(p, M, j). \label{eq:mult_lower_prf}
	\end{align} 
	
	\noindent Finally, using the expression for the mean of a Binomial RV \cite[Ch. 16.2.3.1]{handbook_math}, we write:
	\begin{align}
	\label{eq:expBinom}
		 \sum_{j=1}^M j \cdot v(p, M, j) = \sum_{j=1}^{M} j \cdot {M \choose i} p^{i}(1-p)^{M-i} = Mp.
	\end{align} 
	
	\noindent Combining \eqref{eq:expBinom} with \eqref{eq:mult_lower_prf} and recalling that $p = F_{T_n}(\tau_n)$ we obtain the upper bound in \eqref{eq:capacityUB_mul}.
\end{IEEEproof}	

Next, for asymptotically large $M$, we derive lower and upper bounds on the capacities of systems that use the FA receiver and the average receiver. The derived lower bounds also constitute lower bounds for the capacity expression in \eqref{eq:mTCALNchanCap}, $\mathsf{C}_M(\tau_n)$.


\section{First Arrival and Average Receivers} \label{sec:FA_AVG}
For a large number of released particles, precisely detecting the arrival times of all the particles may become highly complicated. 
This motivates considering two simpler receivers, that can be modeled using additive noise channels. 
The first receiver decodes based on the FA time, whereas the second receiver decodes based on the average arrival time. 
We derive upper and lower bounds on the capacity of a DBMT channel with FA and average detectors, respectively. The derived lower bounds also serve as a lower bound on capacity of the DBMT channel in \eqref{eq:mTCALNchanCap}, which assumes a receiver that can detect the arrival time of all the particles. We now present these receivers and their corresponding channel models.

\subsection{The FA Receiver}

%

Let $\tilde{T}_{n,k} = \min (\vec{T}_{n,k}) $. Then, using the channel model \eqref{eq:mTCALNchan}, the FA receiver applies decoding based on the output of the following channel:
%
\begin{align}
\label{eq:mMinTCALNchan}
\tilde{Y}_k = \begin{cases} \tilde{T}_{y,k}= T_{x,k} + \tilde{T}_{n,k}, & \tilde{T}_{n,k}\leq\tau_n \\ \phi, & \tilde{T}_{n,k}>\tau_n \end{cases}.
\end{align}

\noindent The similarity between the channels \eqref{eq:mMinTCALNchan} and \eqref{eq:TCALNchan_single} is clearly evident, where the difference is only in the difference PDFs of $T_{n,k}$ and $\tilde{T}_{n,k}$. Thus, the capacity of the channel \eqref{eq:mMinTCALNchan} is given by \eqref{eq:TCALNchanCap} (with $T_{n,k}$ replaced by $\tilde{T}_{n,k}$), and lower and upper bounds on this capacity can be obtained using \eqref{eq:capacityLB_single} and \eqref{eq:capacityUB_single}, respectively. To explicitly evaluate these bound we next derive the PDF of $\tilde{T}_{n,k}$.

Clearly, as the channel \eqref{eq:mTCALNchan} is memoryless and i.i.d., $\tilde{T}_{n,k}$ is also i.i.d. for different values of $k$. Thus, in the following we drop the subscript $k$. Using the CDF of $T_n$, the CDF of $\tilde{T}_{n}$ is given by:
\begin{align*}
F_{\tilde{T}_{n}}(t) &= 1-(1-F_{{T}_{n}}(t))^M, 
\end{align*}
and its PDF is given by
\begin{align*}
f_{\tilde{T}_{n}}(t) &= Mf_{{T}_{n}}(t)(1-F_{{T}_{n}}(t))^{M-1}. 
\end{align*}

\noindent Recalling the expressions for the PDF and CDF of the \levy distribution in \eqref{eqn:LevyPDF_0} and \eqref{eqn:LevyCDF}, respectively, calculating the conditional entropy $h(\tilde{T}_n|\tilde{T}_n<\tau_n)$ becomes intractable.
To resolve this issue, we use extreme value theory \cite{ExtremeValueBook} to find the PDF of $\tilde{T}_{n}$ as $M\rightarrow\infty$. We begin with defining the Gumbel distribution:
\begin{definition}[Gumbel Distribution] \label{def:gumbRV}
	Let $\tilde{X}\in\realSet$ be a Gumbel-distributed RV with location parameter $\alpha$ and scale parameter $\beta$. Then, the PDF $\tilde{X}$ is given by:
	\begin{align}
	\label{eq:GumbPDF}
	f_{\tilde{X}}(\tilde{x}) = \tfrac{1}{\beta}\exp\left[\tfrac{\tilde{x}-\alpha}{\beta}-\exp\left(\tfrac{\tilde{x}-\alpha}{\beta}\right) \right], 
	\end{align} 
	and its CDF is given by:
	\begin{align}
	\label{eq:GumbCDF}
	F_{\tilde{X}}(\tilde{x}) = 1-\exp\left[-\exp\left(\tfrac{\tilde{x}-\alpha}{\beta}\right) \right]. 
	\end{align}
	
	\end{definition}
	
	\noindent In the following, we use the notation $\tilde{X} \sim \GumbDist(\alpha,\beta)$ to represent a Gumbel-distributed random variable with parameters $\alpha$ and $\beta$. Having defined the Gumbel distribution, the following lemma presents the distribution of $\tilde{T}_{n}$ for sufficiently large $M$, namely, as $M\rightarrow\infty$.
\begin{lemma}
	\label{lm:MinLevyGumbel}
	Let $\vec{T}_n[i]\sim \LevyDist (0,c)$ be the \ith element of a random delay vector $\vec{T}_n$ of size $M$. Let $\tilde{T}_n = \min(\vec{T}_n)$ be the minimum element of the vector. Then, as $M\rightarrow\infty$, $\tilde{T}_n \sim \GumbDist(\alpha,\beta)$ (i.e., converges to the Gumbel distribution) with the parameters:
	\begin{align}
	\label{eq:paramGumb}
	\alpha = \frac{c}{2\erfcinv^2(\tfrac{1}{M})}, \qquad
	\beta = \alpha-\frac{c}{2\erfcinv^2(\tfrac{1}{Me})}.
	\end{align}
\end{lemma}    
\begin{IEEEproof}
	The proof is provided in Appendix \ref{app:ProofMinLevyGumbel}.
\end{IEEEproof}

Lemma \ref{lm:MinLevyGumbel} facilitates deriving an expression for the conditional entropy $h(\tilde{T}_n|\tilde{T}_n<\tau_n)$ as $M\rightarrow\infty$. To do so, we introduce the following lemma which provides the partial entropy of a Gumbel-distributed RV.
\begin{lemma}
	\label{lm:condGumbEntropy}
	If $\tilde{X}\sim \GumbDist(\alpha,\beta)$, then the partial entropy of $\tilde{X}$ is given by: 
	\begin{align}
	\label{eq:condiEntGumb}
	&\eta(\tilde{X},\tau) \mspace{-3mu}  = \mspace{-3mu}F_{\tilde{X}}(\tau)\log(\beta) 
	 +\log(e)\Bigg[ \mspace{-3mu}\exp\left(\tfrac{\tau-\alpha}{\beta}-\exp(\tfrac{\tau-\alpha}{\beta})\right) \nonumber \\ &  +1\mspace{-3mu}+\mspace{-3mu}\gamma\mspace{-3mu} -\mspace{-3mu}\Ei\left(-e^{\tfrac{\tau-\alpha}{\beta}}\right)
	   +\frac{\tau-\alpha-\beta}{\beta}\exp(-\exp(\tfrac{\tau-\alpha}{\beta}))\Bigg],
	\end{align}
	where $\gamma \mspace{-3mu} \approx \mspace{-3mu} 0.5772$ is the Euler's constant \cite[Ch. 5.2]{nist10}, and $\Ei(\cdot)$ is the exponential integral \cite[Equation 6.2.5]{nist10}.
\end{lemma}   
\begin{IEEEproof}
	The proof is provided in Appendix \ref{app:ProofPartEntGumbel}.
\end{IEEEproof}

\noindent Finally, to find $h(\tilde{T}_n|\tilde{T}_n<\tau_n)$ as $M\rightarrow\infty$ we plug \eqref{eq:GumbPDF} into \eqref{eq:condiEntGumb}, and then plug the resulting expression into \eqref{eqn:timeConstEntropy_3}. Note that the entropy of $\tilde{X}\sim \GumbDist(\alpha,\beta)$ can be obtain from \eqref{eq:condiEntGumb} as
\begin{align}
	\eta(\tilde{X},\tau\rightarrow\infty) = h(\tilde{X})= \log(\beta) + \log(e)(1+\gamma). 
\end{align}

The following theorem provides asymptotic lower and upper bounds on the capacity of \eqref{eq:mMinTCALNchan} using extreme value theory.
\begin{theorem}
\label{th:FA_UBLB}
	The capacity of the FA receiver $\mathsf{C}_M^{\text{FA}}(\tau_n)$, as $M\rightarrow\infty$, is bounded by $\mathsf{C}_M^{\text{FA(lb)}}(\tau_n) \le \mathsf{C}_M^{\text{FA}}(\tau_n) \le\mathsf{C}_M^{\text{FA(ub)}}(\tau_n)$, where the bounds are given by:
	\begin{align}
		\mathsf{C}_M^{\text{FA(lb)}}(\tau_n)& \mspace{-1mu} \triangleq \mspace{-1mu}\underset{\tau_x}{\max}\frac{\left(m(\tau_x,\tau_n,\tilde{T}_n) \mspace{-3mu} - \mspace{-3mu} h(\tilde{T}_n|\tilde{T}_n\leq\tau_n)\right)F_{{\tilde{T}_n}(\tau_n)}}{\tau_x + \tau_n}, \label{eq:capacityLB_FA}\\
		\mathsf{C}_M^{\text{FA(ub)}}(\tau_n)& \mspace{-1mu} \triangleq \mspace{-1mu}\underset{\tau_x}{\max}\frac{\left(\log(\tau_x+\tau_n) \mspace{-3mu} - \mspace{-3mu} h(\tilde{T}_n|\tilde{T}_n\leq\tau_n)\right)F_{{\tilde{T}_n}(\tau_n)}}{\tau_x + \tau_n}, \label{eq:capacityUB_FA}		
	\end{align}
	\noindent with $m(\tau_x,\tau_n,\tilde{T}_n)$ given in \eqref{eqn:mFuncDef}. 
\end{theorem}
\begin{IEEEproof}
	Using the results of Lemma \ref{lm:MinLevyGumbel} and Lemma \ref{lm:condGumbEntropy}, the asymptotic bounds can be derived following same technique used to prove the bounds in Theorem \ref{th:UBLB}.
\end{IEEEproof}

\begin{cor}
	\label{cor:rateFA}
	As $M\rightarrow\infty$, the expression $m(\tau_x,\tau_n,\tilde{T}_n) \mspace{-3mu} - \mspace{-3mu} h(\tilde{T}_n|\tilde{T}_n\leq\tau_n)$ in \eqref{eq:capacityLB_FA} scales at least as $\log(\log(M))$.  
\end{cor}
\begin{IEEEproof}
	The proof is provided in Appendix \ref{app:ProofScallingFA}.
\end{IEEEproof}
\begin{rem}
	Although the numerator of \eqref{eq:capacityLB_FA} scales as $\log(\log(M))$, as is shown in Section \ref{sec:NumEvaluation}, the capacity may scale faster. This follows as by increasing the number of particles, the optimal $\tau_x$ and $\tau_n$ values decrease.
\end{rem}

Next, we derive asymptotic lower and upper bounds on the capacity of the average receiver.

\subsection{The Average Receiver}

The average receiver applies decoding based on the average arrival times during each channel use. 
This is equivalent to decoding based on the output of the following channel:
%
\begin{align}
\label{eq:mTCALNchanAVG}
\vec{Y}_k \mspace{-3mu} = \mspace{-3mu}
\begin{cases}
\tfrac{1}{|\mathcal{J}_k|}\underset{i\in\mathcal{J}_k}{\sum}\vec{T}_{y,k}[i] \mspace{-3mu} = \mspace{-3mu} T_{x,k} \mspace{-3mu} + \mspace{-3mu} \tfrac{1}{|\mathcal{J}_k|}\underset{i\in\mathcal{J}_k}{\sum}\vec{T}_{n,k}[i], & |\mathcal{J}_k| \mspace{-3mu} > \mspace{-3mu} 0 \\
\phi, &  |\mathcal{J}_k| \mspace{-3mu} = \mspace{-3mu} 0
\end{cases}.
\end{align}

To derive the asymptotic bounds on the capacity of this channel, as a function of system parameters, we formally define the {\em truncated \levy distribution}, i.e., the distribution of $T_n$ given $T_n<\tau_n$, and its corresponding first and second moments. 
\begin{definition}[Truncated \levy Distribution] \label{def:trunLevyRV}
	Let $X$ be a truncated \levy distribution with parameters $0<c<\infty$ and $0<\tau<\infty$. Then the PDF of $X$ is given by:
	\begin{align}
		\label{eq:trunLevyPDF}
		f_X(x;c,\tau) = \begin{cases}
		\erfc^{-1}\mspace{-3mu}\left(\sqrt{\frac{c}{2\tau}}\right)\mspace{-3mu}\sqrt{\frac{c}{2 \pi x^3}}\exp \left( -\frac{c}{2x} \right), &\mspace{-15mu} 0<x\leq\tau \\
		0, & \mspace{-15mu} \text{otherwise}
		\end{cases},
	\end{align}
	
	\noindent and	the first and second moments of $X$ are given by:\footnote{These moments were calculated using {\tt Mathematica}.}
	\begin{align}
	\mathbb{E}[X] \mspace{-3mu} &= \mspace{-3mu} \frac{1}{\erfc(\sqrt{\tfrac{c}{2\tau}})}\left[\sqrt{\tfrac{2c\tau}{\pi}}~_1F_1[-\tfrac{1}{2},\tfrac{1}{2},-\tfrac{c}{2\tau}] - c\right], \label{eq:m1TrunLevy}\\
		\mathbb{E}[X^2] \mspace{-3mu} &= \mspace{-3mu} \frac{1}{3\erfc(\sqrt{\tfrac{c}{2\tau}})}\left[\sqrt{\tfrac{2c\tau^3}{\pi}}~_1F_1[-\tfrac{3}{2},-\tfrac{1}{2},-\tfrac{c}{2\tau}] \mspace{-3mu} + \mspace{-3mu} c^2\right].\label{eq:m2TrunLevy}
	\end{align}
\end{definition}
\noindent The variance of a truncated \levy RV can be calculated using its first and the second moments. Note that although the mean and the variance of a \levy RV is infinite, the mean and the variance of a truncated \levy RV are finite.

The following lemma characterizes the asymptotic behavior of the channel in \eqref{eq:mTCALNchanAVG}.
\begin{lemma}
	\label{lm:convToGausChan}
	As $M\rightarrow\infty$, the channel in \eqref{eq:mTCALNchanAVG} converges to an equivalent channel given by:
	\begin{align}
	\label{eq:mTCALNchanGaus}
	\hat{T}_y = T_x + \hat{T}_n,
	\end{align} 
	where $\hat{T}_n\sim\NormDist(0,\frac{\var[T^\prime_{n}]}{M F_{T_n}(\tau_n)})$ is an additive Gaussian noise, $T^\prime_{n}$ is a truncated \levy RV with parameters $c$ and $\tau_n$, and $\hat{T}_y$ is the channel output.	
\end{lemma}
\begin{IEEEproof}
	The proof is provided in Appendix \ref{app:ProofLemmaConvGausChan}.
\end{IEEEproof}

The following theorem presents asymptotic lower and upper bounds on the capacity of the channel in \eqref{eq:mTCALNchanAVG}.
\begin{theorem}
\label{th:AVG_UBLB}
	The capacity of the average receiver $\mathsf{C}_M^{\text{AV}}(\tau_n)$, as $M\rightarrow\infty$, is bounded by $\mathsf{C}_M^{\text{AV(lb)}}(\tau_n) \le \mathsf{C}_M^{\text{AV}}(\tau_n) \le \mathsf{C}_M^{\text{AV(ub)}}(\tau_n)$, where the bounds are given by:
	\begin{align}
	\mathsf{C}_M^{\text{AV(lb)}}(\tau_n)& \mspace{-1mu} \triangleq \mspace{-1mu}\underset{\tau_x}{\max}\frac{0.5 \log \big( \tau_x^2+2^{2h(\hat{T}_n)}\big) \mspace{-3mu} - \mspace{-3mu} h(\hat{T}_n)}{\tau_x + \tau_n}, \label{eq:capacityLB_Avg} \\
	\mathsf{C}_M^{\text{AV(ub)}}(\tau_n)& \mspace{-1mu} \triangleq \mspace{-1mu}\underset{\tau_x}{\max}\frac{\log \big( \tau_x+\tau_n\big) \mspace{-3mu} - \mspace{-3mu} h(\hat{T}_n)}{\tau_x + \tau_n}, \label{eq:capacityUB_Avg}
	\end{align}
	\noindent where $\hat{T}_n$ is given in Lemma \ref{lm:convToGausChan} and 
	\begin{align}
		h(\hat{T}_n) = \tfrac{1}{2} \log\left(2\pi e \frac{\var[T^\prime_{n}]}{M F_{T_n}(\tau_n)}\right). \label{eq:entrGaus}
	\end{align}. 
\end{theorem}
\begin{IEEEproof}
	Using Lemma \ref{lm:convToGausChan}, the asymptotic bounds can be derived following the same technique used to prove the bounds in Theorem \ref{th:UBLB}.
\end{IEEEproof}
\begin{cor}
	\label{cor:rateAvg}
	As $M\rightarrow\infty$, the numerator of the right-hand-side of \eqref{eq:capacityLB_Avg} scales at least as $\log(M)$.  
\end{cor}
\begin{IEEEproof}
	The proof follows directly from the fact that $\var[T^\prime_{n}]$ and $F_{T_n}(\tau_n)$ are bounded.
\end{IEEEproof}

\begin{rem}
	Based on the results of the Corollaries \ref{cor:rateFA} and \ref{cor:rateAvg}, one might suspect that {\em asymptotically} an average receiver is universally better than a FA receiver. 
	However, this strongly depends on the {\em distribution} of the additive noise. 
	For example, if the additive noise is uniformly distributed (instead of a truncated L\'evy), it can be shown that the first arrival receiver can achieve higher information rates compared to the average receiver. 
\end{rem}

Next, we numerically evaluate our bounds on the capacity of the DBMT channel.

\section{Numerical Results}
\label{sec:NumEvaluation}

\begin{figure*}[t!]
	\normalsize
	\centering
	\begin{minipage}{.43\textwidth}
		\begin{center}
			\includegraphics[width=\columnwidth,keepaspectratio]{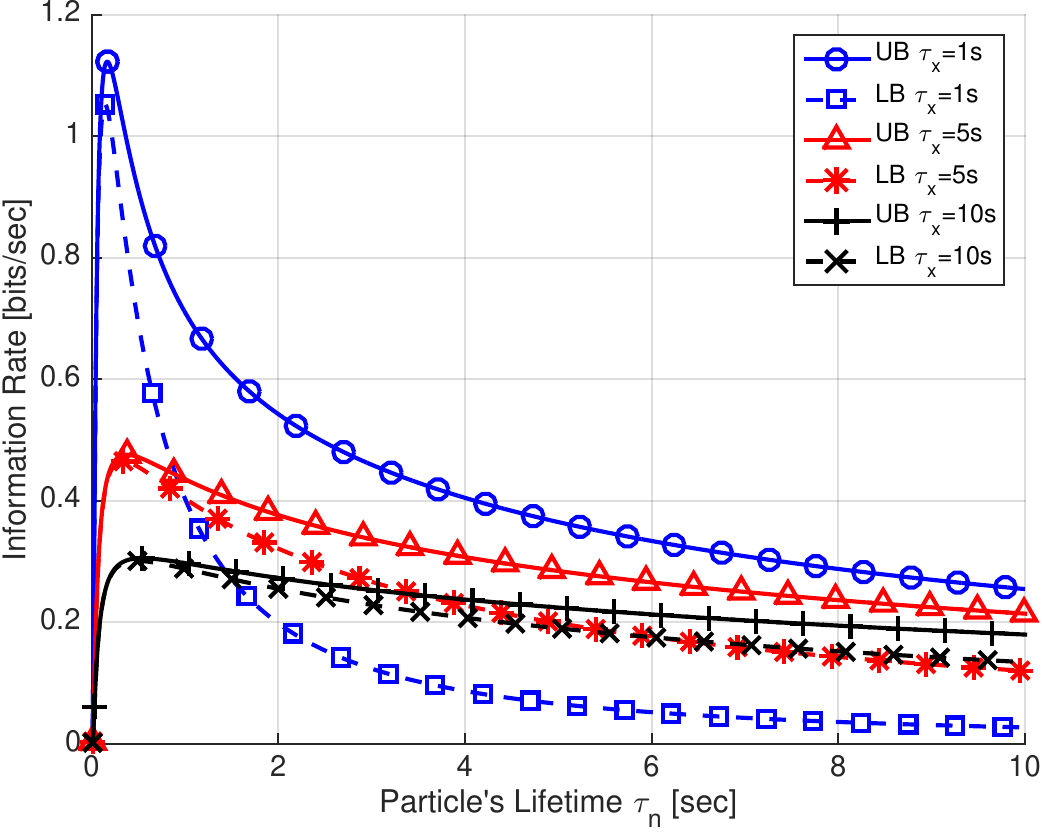}
		\end{center}
		\vspace{-0.35cm}
		\caption{\label{fig:boundsPUC_VStnC0p1} $\mathsf{C}_1^{\text{lb}}(\tau_n)$ and $\mathsf{C}_1^{\text{ub}}(\tau_n)$ versus the particle's lifetime $\tau_n$, for $\tau_x=1, 5, 10$ [sec], and $c = 0.1$.}
	\end{minipage}
	\hspace{0.8cm}
	\begin{minipage}{.43\textwidth}
		\begin{center}
			\includegraphics[width=\columnwidth,keepaspectratio]{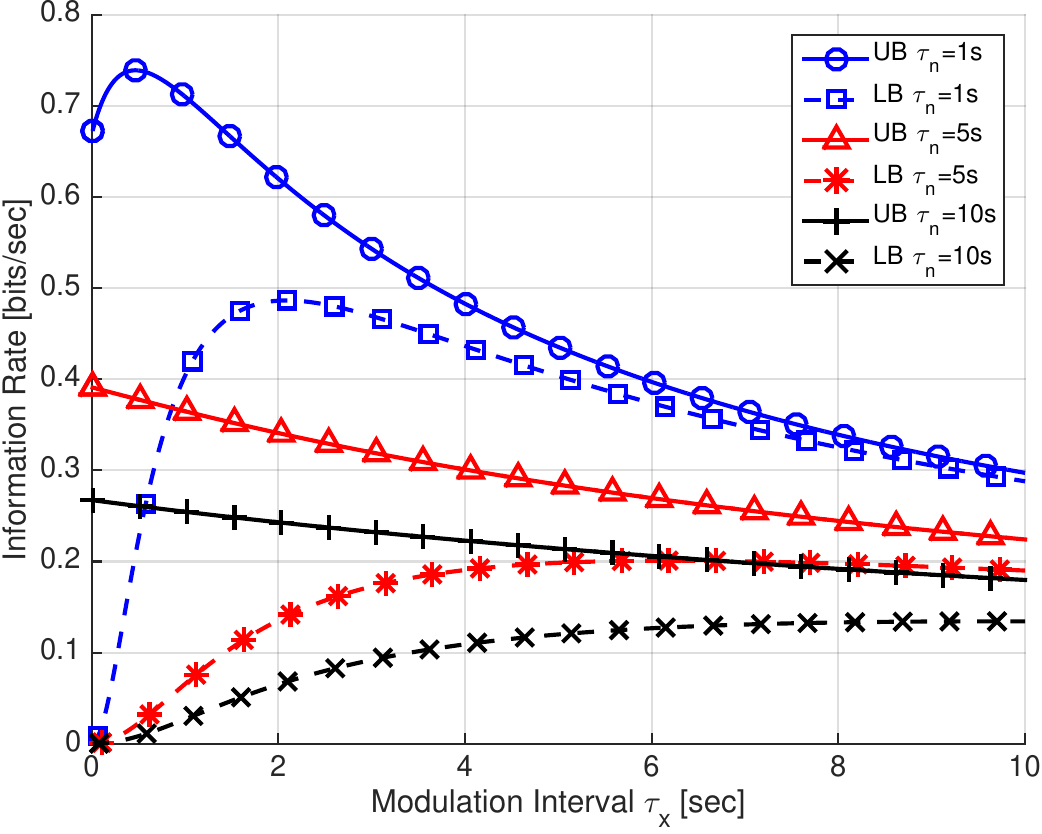}
		\end{center}
		\vspace{-0.35cm}
		\caption{\label{fig:boundsPUC_VStxC0p1} $\mathsf{C}^{\text{lb}}_1(\tau_x, \tau_n)$ and $\mathsf{C}^{\text{ub}}_1(\tau_x, \tau_n)$ versus the symbol interval $\tau_x$, for $\tau_n=1, 5, 10$ [sec], and $c = 0.1$.}
	\end{minipage}
\end{figure*}

\begin{figure}
	\begin{center}
		\includegraphics[width=1\columnwidth,keepaspectratio]{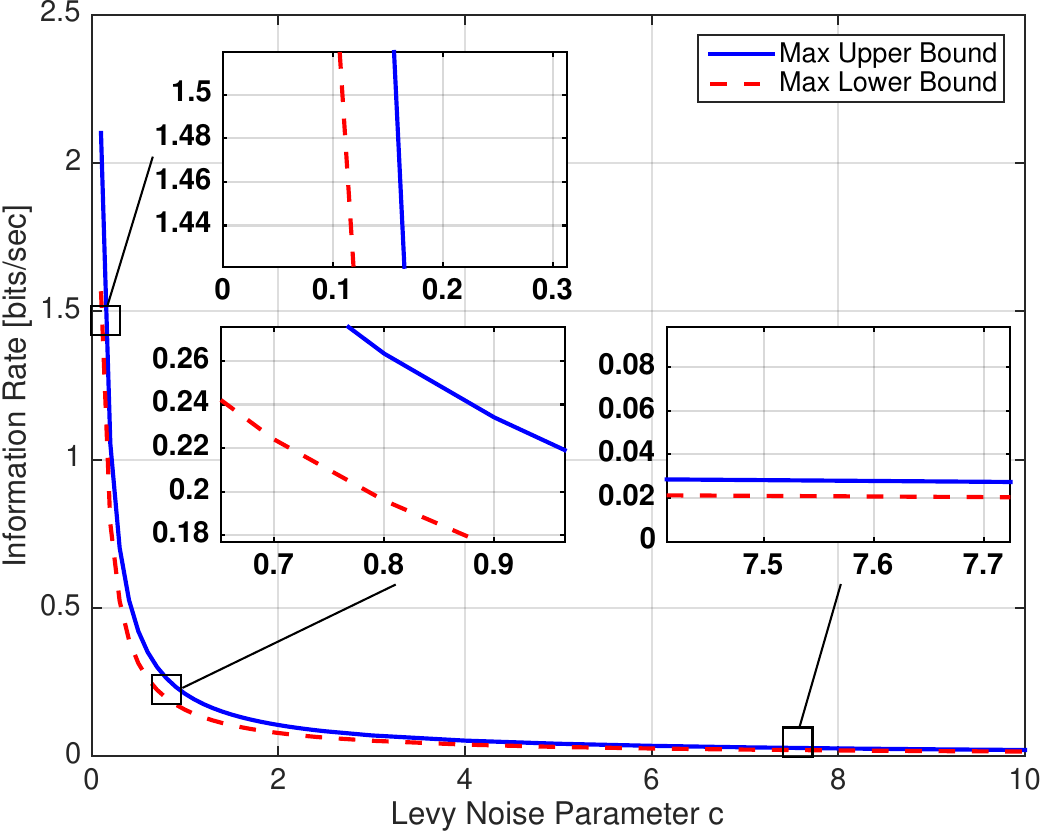}
	\end{center}
	\caption{\label{fig:maxBoundsVScn} The maximum lower and upper bounds on the capacity versus the \levy noise parameter $c$. The lower and upper bounds are simultaneously maximized over $\tau_n$ and $\tau_x$.}
	\vspace{0.2cm}
\end{figure}

\begin{figure*}
	\normalsize
	\centering
	\begin{minipage}{.43\textwidth}
		\begin{center}
			\includegraphics[width=\columnwidth,keepaspectratio]{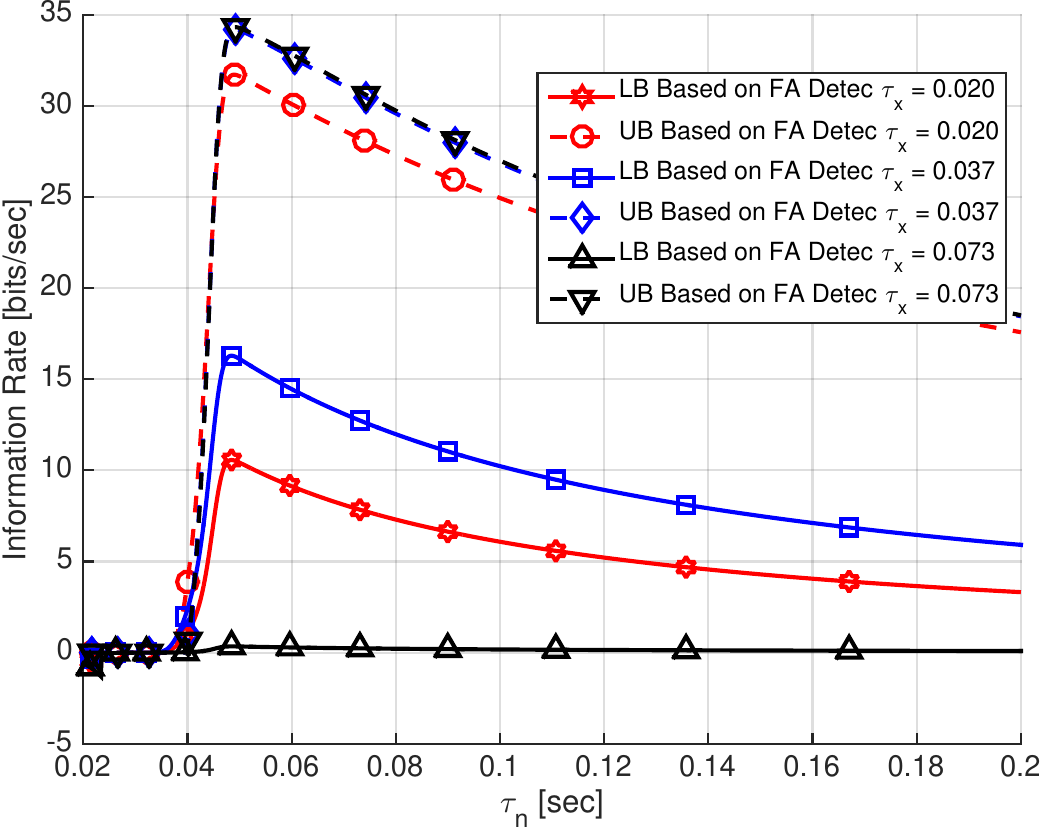}
		\end{center}
		\vspace{-0.35cm}
		\caption{\label{fig:boundsFA_VStnC1} $\mathsf{C}_{M}^{\text{FA(lb)}}(\tau_n)$ and $\mathsf{C}_{M}^{\text{FA(lb)}}(\tau_n)$ versus the particle's lifetime $\tau_n$, for $\tau_x=0.02, 0.037, 0.073$ [sec], $M=10^6$, and $c = 1$.}
	\end{minipage}
	\hspace{0.6cm}
	\begin{minipage}{.43\textwidth}
		\begin{center}
			\includegraphics[width=\columnwidth,keepaspectratio]{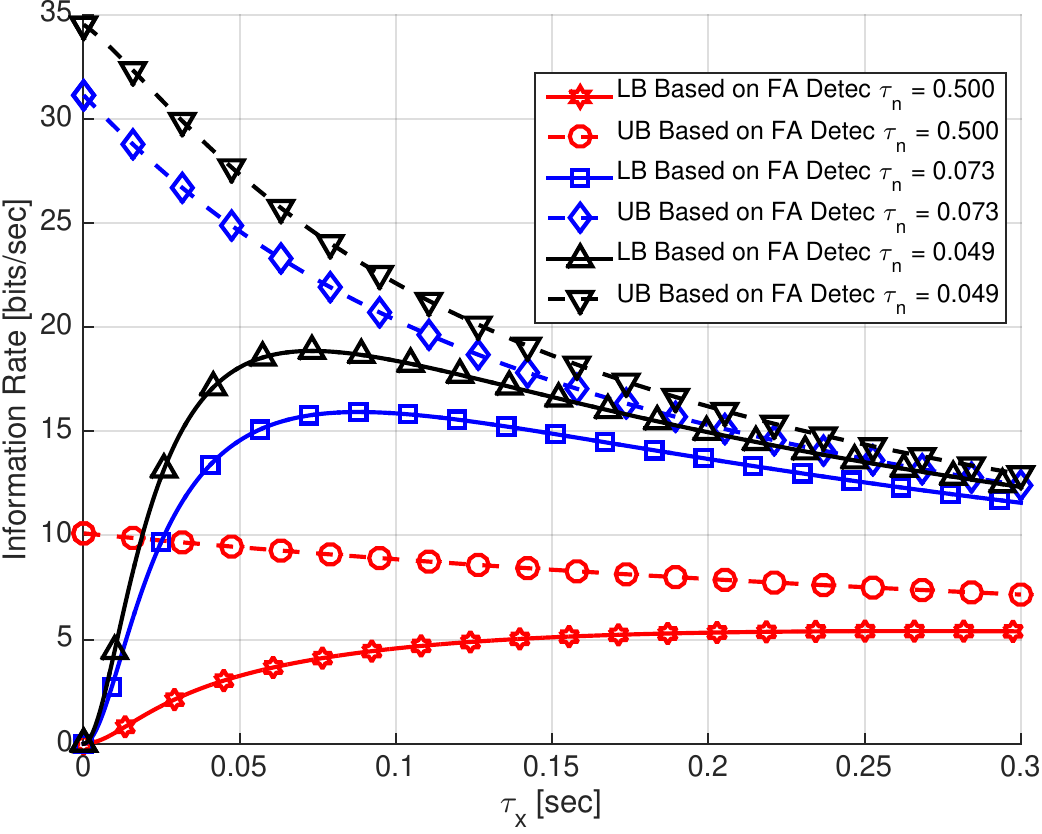}
		\end{center}
		\vspace{-0.35cm}
		\caption{\label{fig:boundsFA_VStxC1} $\mathsf{C}^{\text{FA(lb)}}_{M}(\tau_x, \tau_n)$ and $\mathsf{C}^{\text{FA(ub)}}_{M}(\tau_x, \tau_n)$ versus the symbol interval $\tau_x$, for $\tau_n=0.049, 0.073, 0.5$ [sec], $M=10^6$, and $c = 1$.}
	\end{minipage}
\end{figure*}
\begin{figure*}
	\normalsize
	\centering
	\begin{minipage}{.43\textwidth}
		\begin{center}
			\includegraphics[width=\columnwidth,keepaspectratio]{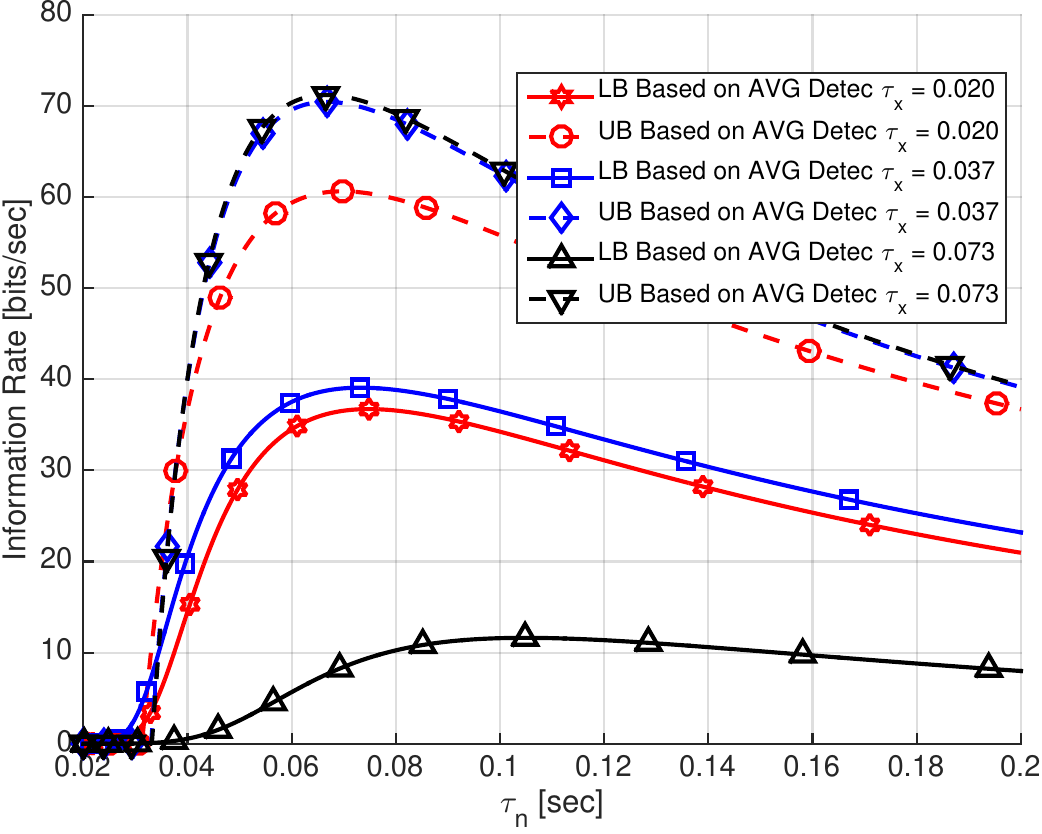}
		\end{center}
		\vspace{-0.35cm}
		\caption{\label{fig:boundsAVG_VStnC1} $\mathsf{C}_{M}^{\text{AV(lb)}}(\tau_n)$ and $\mathsf{C}_{M}^{\text{AV(lb)}}(\tau_n)$ versus the particle's lifetime $\tau_n$, for $\tau_x=0.02, 0.037, 0.073$ [sec], $M=10^6$, and $c = 1$.}
	\end{minipage}
	\hspace{0.6cm}
	\begin{minipage}{.43\textwidth}
		\begin{center}
			\includegraphics[width=\columnwidth,keepaspectratio]{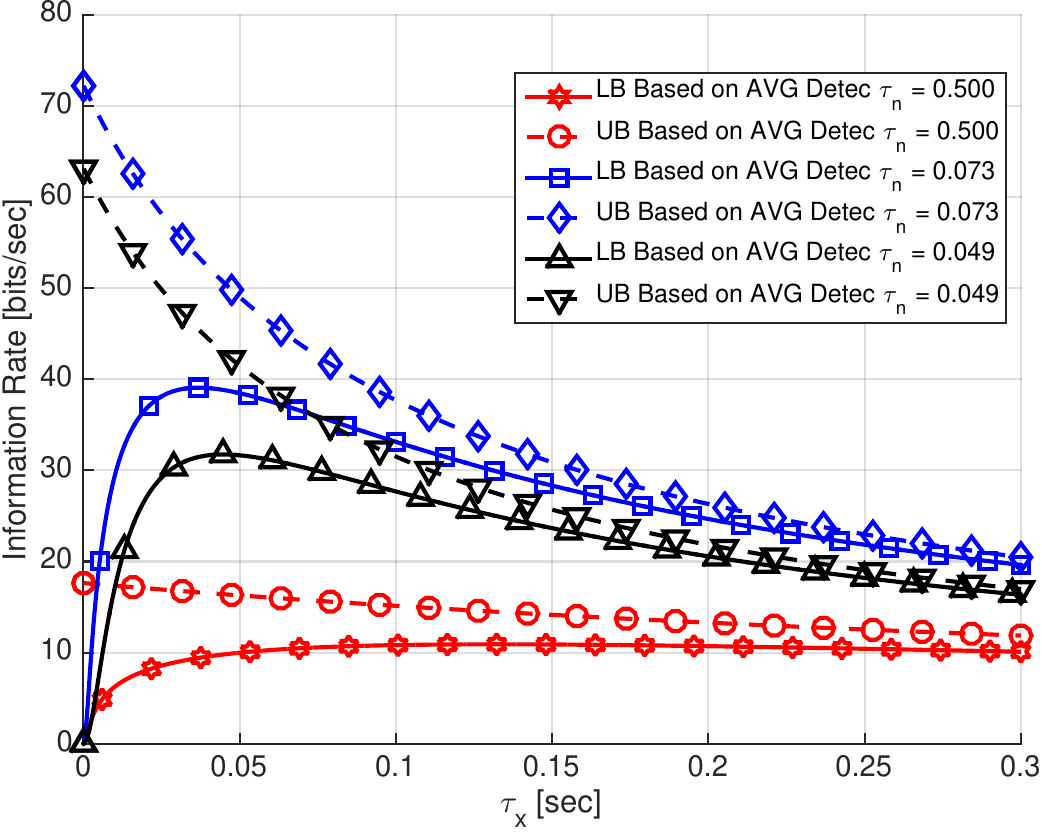}
		\end{center}
		\vspace{-0.35cm}
		\caption{\label{fig:boundsAVG_VStxC1} $\mathsf{C}^{\text{AV(lb)}}_{M}(\tau_x, \tau_n)$ and $\mathsf{C}^{\text{AV(ub)}}_{M}(\tau_x, \tau_n)$ versus the symbol interval $\tau_x$, for $\tau_n=0.049, 0.073, 0.5$ [sec], $M=10^6$, and $c = 1$.}
	\end{minipage}
\end{figure*}

We begin our numerical evaluations with the lower and upper bounds on the capacity of the single-particle DBMT channel. Note that the capacity in \eqref{eq:TCALNchanCap} and the corresponding lower and upper bounds in \eqref{eq:capacityLB_single}--\eqref{eq:capacityUB_single} depend on three system parameters other than the input distribution: the symbol interval $\tau_x$, the particle's lifetime $\tau_n$, and the \levy noise parameter $c$, which is a function of the distance between the transmitter and the receiver and the diffusion coefficient. In this section, the effect of each parameter on the channel capacity is investigated by numerically evaluating the upper and lower bounds in different scenarios. 

Fig. \ref{fig:boundsPUC_VStnC0p1} depicts the arguments of the maximization problems \eqref{eq:capacityLB_single} and \eqref{eq:capacityUB_single}, for the single particle DBMT channel, with respect to particle's lifetime $\tau_n$ for $\tau_x=1,5,10$ [sec], and for $c = 0.1$. As can be seen from the plots, the lower and upper bounds are tight for small values of $\tau_n$ and diverge as $\tau_n$ increases. This follows as $h(T_n | T_n \le \tau_n) \to -\infty$ when $\tau_n \to 0$.\footnote{Recall that $h(T_n | T_n \le \tau_n) \le \log(\tau_n) \to -\infty$, when $\tau_n \to 0$.} 
It can further be noted that although the bounds are not tight as $\tau_n$ increases, they are tight before the peaks. 
Based on these results, an interesting and nontrivial observation is that the $\tau_n$ which maximizes the capacity (given a fixed $\tau_x$) tends to be small. Therefore, it is best to use information particles that have a short lifetime and quickly degrade in the environment after they are released.

In Fig. \ref{fig:boundsPUC_VStxC0p1} we investigate the effect of the symbol interval on channel capacity by plotting the bounds on capacity versus $\tau_x$, for $\tau_n=1,5,10$ [sec], and for  $c = 0.1$. As the values of $\tau_x$ tends to zero, the bounds are not tight, while as $\tau_x$ increases they converge, as stated in Remark \ref{rem:argumentsConverge}. For smaller particle's lifetime $\tau_n$, the bounds tend to converge more rapidly. 
Note that in Fig. \ref{fig:boundsPUC_VStxC0p1}, for a given $\tau_n$, the lower and upper bounds are maximized by different values of $\tau_x$. Therefore, it is not clear from the plots which value of $\tau_x$ maximizes the capacity. However, it can be observed that, similarly to Fig. \ref{fig:boundsPUC_VStnC0p1}, the bounds achieve their maximal values for relatively small values of $\tau_x$. 

\begin{figure*}
	\normalsize
	\centering
	\begin{minipage}{.43\textwidth}
		\begin{center}
			\includegraphics[width=\columnwidth,keepaspectratio]{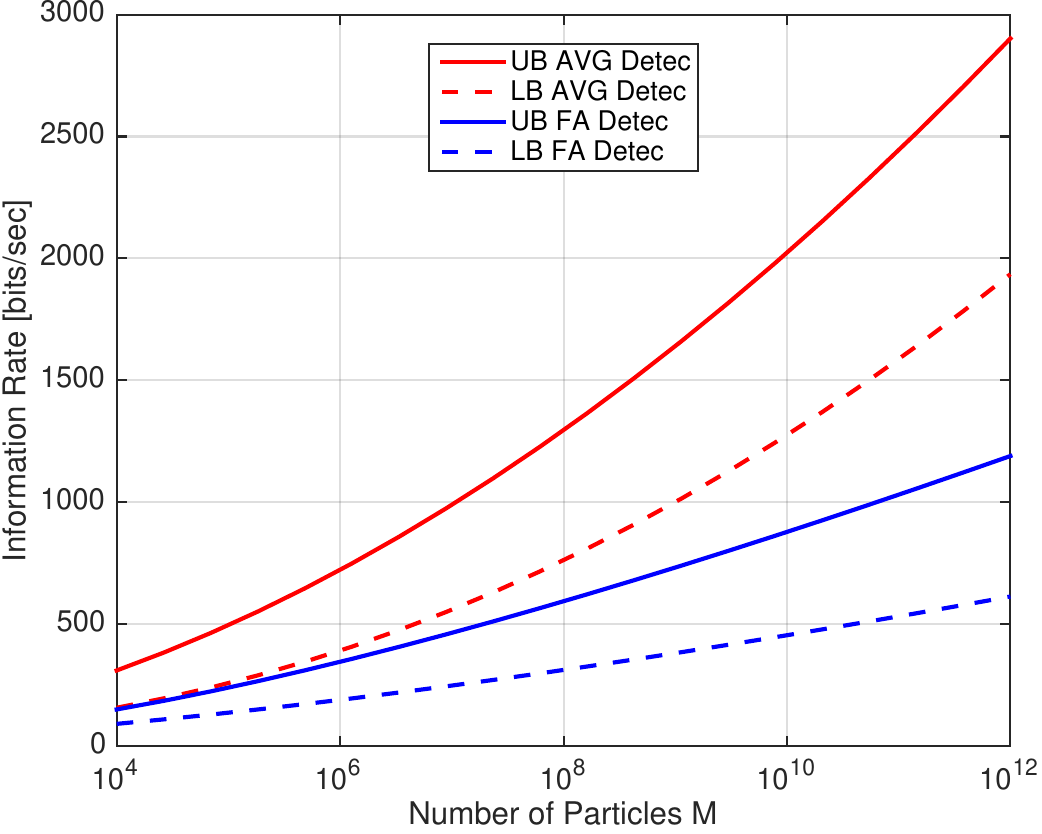}
		\end{center}
		\vspace{-0.35cm}
		\caption{\label{fig:bounds_VS_M_C0p1} The maximum asymptotic lower and upper bounds on the capacity of the DBMT channel with FA and average receivers for $c = 0.1$. The lower and upper bounds are simultaneously maximized over $\tau_n$ and $\tau_x$.}
	\end{minipage}
	\hspace{0.6cm}
	\begin{minipage}{.43\textwidth}
		\begin{center}
			\includegraphics[width=\columnwidth,keepaspectratio]{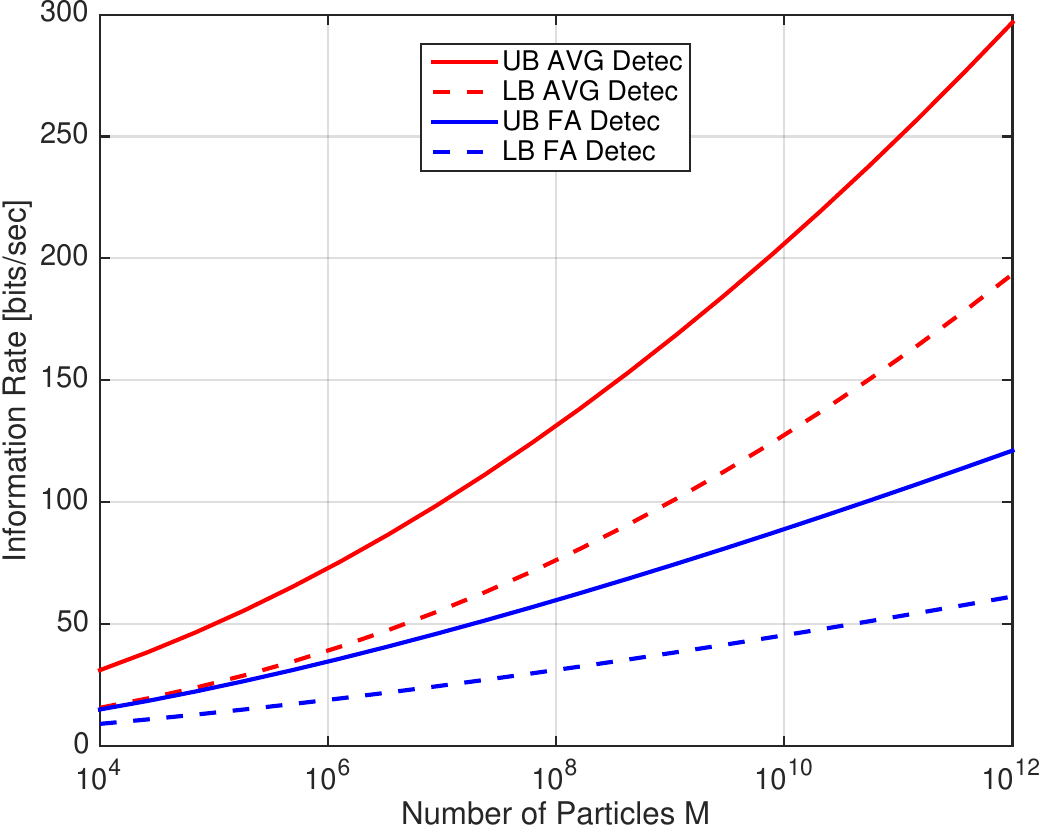}
		\end{center}
		\vspace{-0.35cm}
		\caption{\label{fig:bounds_VS_M_C1} The maximum asymptotic lower and upper bounds on the capacity of the DBMT channel with FA and average receivers for $c = 1$. The lower and upper bounds are simultaneously maximized over $\tau_n$ and $\tau_x$.}
	\end{minipage}
\end{figure*}

Next, we study the effect of the \levy noise parameter $c$ on the capacity of the single-particle DBMT channel. For this purpose, we numerically maximize the lower and upper bounds on the capacity with respect to $\tau_x$ and $\tau_n$. Note that by using the maximizing $\tau_x$ and $\tau_n$ one maximizes the information rate (in bits per second) of the considered communication system. 
Fig.~\ref{fig:maxBoundsVScn} depicts the maximal lower and upper bounds as a function of $c$. The maximizing $\tau_x$ and $\tau_n$ are detailed in Table~\ref{tab:maximizingTaus}.
\begin{table}[h]
	\begin{center}
		\begin{tabular}[t]{|c|c|c|c|c|c|c|}
			\hline
			$c$ & 0.1  & 0.5 & 1 & 2 & 4 & 8 \\
			\hline
			\hline
			$ \tau_x^{\text{lb}}$  &0.17  &0.8  &1.63  &3.26 &6.52 &13.04  \\
			\hline
			$ \tau_n^{\text{lb}} $ &0.06  &0.29  &0.59  &1.18 &2.36 &4.72  \\
			\hline
			$ \tau_x^{\text{ub}}$  &0.06  &0.32  &0.65  & 1.31&2.61 &5.24  \\
			\hline
			$ \tau_n^{\text{ub}} $ &0.05  & 0.27 &0.54  &1.09 & 2.17&4.35  \\
			\hline
		\end{tabular}
		\vspace{0.3cm}
		\caption{The maximizing values of $\tau_x$ and $\tau_n$, for the lower and upper bounds in \eqref{eq:capacityLB_single}--\eqref{eq:capacityUB_single}, for different values of $c$. \label{tab:maximizingTaus}}
		\vspace{-0.5cm}
	\end{center}
\end{table}
 It can be observed that the capacity drops exponentially with respect to $c$.  The increase in $c$ can result from either an increase in the distance between the transmitter and the receiver, or a decrease in the diffusion coefficient of the information particles with respect to the propagation medium. To provide an example, the diffusion coefficient for glucose in water at 25$^\circ$C is 600 $\mu$m$^2$/s \cite{far16ST}. Therefore, if the separation distance between the transmitter and receiver is 10 $\mu$m, the \levy noise parameter is $c=0.083$, and if the separation distance is 50 $\mu$m, the \levy noise parameter is $c= 2.083$. If glycerol is used instead of glucose, the diffusion coefficient would change to 930 $\mu$m$^2$/s \cite{far16ST}, and the \levy noise parameters would be $c=0.054$ and $c=1.344$, respectively.

Finally, we consider the DBMT channel with diversity where the number of particles $M$ is large, and study the bounds on capacity of systems equipped with the FA receiver and the average receiver. 
Figs.~\ref{fig:boundsFA_VStnC1} and \ref{fig:boundsFA_VStxC1} show how the asymptotic bounds on the capacity of the FA receiver change with $\tau_n$ and $\tau_x$, respectively, while Figs.~\ref{fig:boundsAVG_VStnC1} and \ref{fig:boundsAVG_VStxC1} shows the same for the average receiver. In these plots the number of particles is fixed to $M=10^6$, and the \levy noise parameter is $c=1$. As can be seen from the plots, the bounds behave similarly to those in Figs.~\ref{fig:boundsPUC_VStnC0p1} and \ref{fig:boundsPUC_VStxC0p1}. 

Figs.~\ref{fig:bounds_VS_M_C0p1} and \ref{fig:bounds_VS_M_C1} shows the scaling behavior of lower and upper bounds on capacity for the FA and average receivers. In Fig. \ref{fig:bounds_VS_M_C0p1} the \levy noise parameter is $c=0.1$, while in \ref{fig:bounds_VS_M_C1} it is $c=1$. We emphasize that in these plots the bounds are simultaneously maximized over $\tau_n$ and $\tau_x$. Both plots support the results of Corollaries \ref{cor:rateFA} and \ref{cor:rateAvg} indicating that the average receiver can achieve higher information rates compared to the FA receiver. In this context, one should remember that this is a result of the specific nature of the truncated \levy distribution, and for noise distributions such as uniform, FA can achieve higher rates. By using the maximizing particle lifetime, the rates increase is {\em polylogarithmic} with the number of particles. In fact, using basic curve fitting techniques, it can be shown that each plot can be represented using a quadratic equation. This follows since by increasing the number of particles, one can use particles with a shorter lifetime, which results in an increase in rate in bits per second.  

We conclude this section with noting that the lower bounds presented in \eqref{eq:capacityLB_FA} and \eqref{eq:capacityLB_Avg} are derived for {\em sufficiently large $M$}. These bounds scale at most {\em poly-logarithmically} with $M$, while the upper bound in \eqref{eq:capacityUB_mul} scales linearly. Therefore, jointly plotting these lower and upper bound is not informative.


\vspace{-0.2cm}
\section{Conclusions and Future Work}
\label{sec:concl}

In this work we considered MT channels, where information is modulated on the release time of particles, and showed that these channels can be represented as an additive noise channel. By assuming that the information particles have a finite lifetime, we formally defined the capacity of the MT channels. We then showed that the \levy distribution can be used to formulate the DBMT channel, and derived upper and lower bounds on capacity of this channel. We further showed that by simultaneously releasing multiple particles, the capacity increases, which is analogous to receiver diversity as each particle propagates to the receiver independently. We further showed that this increase in capacity is at least poly-logarithmic with respect to the number of information particles released. 

Finally, we numerically evaluated the upper and lower bounds on capacity, for both the single-particle DBMT channel and the DBMT channel with diversity, and analytically showed that our bounds converge for large symbol durations. Moreover, the bounds are tight when they are simultaneously maximized over both the symbol interval and the particle's lifetime. The maximizing particle lifetime was observed to be small (within several seconds). This implies that it is better to quickly remove the information particles in the channel with different techniques such as using enzymes or chemical reactions. Similarly, the numerical evaluations indicate that the bounds are maximized when the symbol interval (i.e. the time period where the transmitter could encode a message by releasing particles) is short, i.e., within few tens of seconds.

As part of future work, it is desirable to incorporate more realistic degradation models for information particles, such as the exponential distribution, which well-models the exponential decay of particles. Another research direction is to compare and combine the capacity expressions between the MT channels, and concentration-based channels, where the information is encoded on the number of particles. Extending the results to channels with memory is another important area of future work. 

\appendices

\vspace{-0.3cm}
\section{Proof of \eqref{eq:capacityDef}} \label{annex:BasicCapacityProof}

\subsection{Achievability}
We show that for every rate $R < \mathsf{C}(\tau_n)$, there exists a sequence of $(K, R, \tau_x, \tau_n)$ codes with average probability of error $P_e^{(K)}$ that tends to zero as $K \to \infty$. For simplicity, we assume that $2^{K(\tau_x + \tau_n)R}$ is an integer.

\subsubsection{Codebook Construction}

Fix a density $f_{T_{x}}(T_x) \in \mathcal{F}(\tau_x)$. 
Generate $2^{K(\tau_x + \tau_n)R}$ sequences $\{T_{x,k}\}_{k=1}^K(w), w \in \mathcal{W}$, by choosing the letters $T_{x,k}(w)$ independently according to the density $f_{T_{x}}(T_x - (k-1) \cdot (\tau_x + \tau_n))$, namely, $T_{x,k}(w) \in \mathcal{A}_k$. 
Next, we follow the approach of \cite[pgs. 251--252]{cover-book} and let $\mathcal{P}$ be a partition of $\mathcal{A}_1$, i.e., $\mathcal{P}$ is a finite collection of disjoint sets $\mathcal{P}_i$ such that $\cup_i \mathcal{P}_i = \mathcal{A}_1$. 
We further let $[T_{x,1}]_{\mathcal{P}}$ denote the quantization of $T_{x,1}$ by $\mathcal{P}$. 
Similarly, by noting that $f_{T_x}(T_{x,k}) = f_{T_{x}}(T_x - (k-1) \cdot (\tau_x + \tau_n))$, we define $[T_{x,k}]_{\mathcal{P}}$. The sequences $\{[T_{x,k}]_{\mathcal{P}}\}_{k=1}^K(w)$ constitute the codebook $\mathcal{C}$, which is known to both the transmitter and receiver. 

\subsubsection{Encoding}

To send the message $w \in \mathcal{W}$, the transmitter sends $\{[T_{x,k}]_{\mathcal{P}}\}_{k=1}^K(w)$.

\subsubsection{Decoding}
Let $\mathcal{Q}$ be a quantization of the outputs $Y_k$, defined in the same manner as $\mathcal{P}$. 
The receiver declares that $\hat{W} \in \mathcal{W}$ is sent if it is the unique message such that $\left(\{[T_{x,k}]_{\mathcal{P}}\}_{k=1}^K(\hat{w}), \{[Y_{k}]_{\mathcal{Q}}\}_{k=1}^K \right) \in \styp([T_x]_{\mathcal{P}}, [Y]_{\mathcal{Q}}), Y_{k} \in \mathcal{B}_k$. If no such $w$ exists, the receiver declares an error.

\subsubsection{Error Probability Analysis}
As the messages are uniformly distributed over $\mathcal{W}$, and from the symmetry of the random codebook construction, we assume without loss of generality that $W=1$ was sent. The receiver makes an error if and only if one or both of the following events occur:
\begin{align*}
	\mathcal{E}_1 & \mspace{-3mu} \triangleq \mspace{-3mu} \left\{\left(\{[T_{x,k}]_{\mathcal{P}}\}_{k=1}^K(1), \{[Y_{k}]_{\mathcal{Q}}\}_{k=1}^K \right) \mspace{-2mu} \notin \mspace{-2mu} \styp \right\} \\
	\mathcal{E}_2 & \mspace{-3mu} \triangleq \mspace{-3mu} \Big\{ \exists \tilde{w} \in \mathcal{W} \mspace{-2mu}: \mspace{-2mu} \tilde{w} \neq 1, \\
	& \mspace{40mu} \left(\{[T_{x,k}]_{\mathcal{P}}\}_{k=1}^K(\tilde{w}), \{[Y_{k}]_{\mathcal{Q}}\}_{k=1}^K \right) \mspace{-2mu} \in \mspace{-2mu} \styp \Big\}.
\end{align*}

\noindent Thus, by the union bound $P_e^{(K)} = \Pr \{\mathcal{E}_1 \cup \mathcal{E}_2 | W=1 \} \le \Pr \{\mathcal{E}_1 | W = 1\} + \Pr \{\mathcal{E}_2 | W=1\}$. Now, from \cite[Lemma 10.6.1]{cover-book} it follows that $\Pr \{\mathcal{E}_1 | W = 1\} \to 0$ as $K \to \infty$, and therefore $\Pr \{\mathcal{E}_1 | W = 1\} \le \epsilon$ for sufficiently large $K$. Next, we note that for $\tilde{w} \neq 1$ we have $\{T_{x,k}\}_{k=1}^K(\tilde{w})$ and $\{Y_{k}\}_{k=1}^K$ independent, and therefore $\{[T_{x,k}]_{\mathcal{P}}\}_{k=1}^K(\tilde{w})$ and $\{[Y_{k}]_{\mathcal{Q}}\}_{k=1}^K$ are also independent. Hence, from \cite[Lemma 10.6.2]{cover-book} we have: 
\vspace{-0.2cm}
\begin{align*}
	& \Pr \left\{ \left(\{[T_{x,k}]_{\mathcal{P}}\}_{k=1}^K(\tilde{w}), \{[Y_{k}]_{\mathcal{Q}}\}_{k=1}^K \right) \mspace{-2mu} \in \mspace{-2mu} \styp | W \mspace{-3mu} = \mspace{-3mu} 1 \right\} \\
	& \mspace{250mu} \le 2^{-K(I([T_x]_{\mathcal{P}};[Y]_{\mathcal{Q}}) - \epsilon_1)},
\end{align*}

\vspace{-0.15cm}
\noindent where $\epsilon_1 \to 0$ as $\epsilon \to 0$ and $K \to \infty$. Again, using the union bound, we write:
\vspace{-0.2cm}
\begin{align*}
	& \Pr \{\mathcal{E}_2 | W=1\} \\
	& \mspace{8mu} \le \mspace{-20mu} \sum_{w=2}^{2^{K(\tau_x + \tau_n)R}} \mspace{-25mu} \Pr \left\{ \mspace{-2mu} \left(\{[T_{x,k}]_{\mathcal{P}}\}_{k=1}^K(\tilde{w}), \{[Y_{k}]_{\mathcal{Q}}\}_{k=1}^K \right) \mspace{-2mu} \in \mspace{-2mu} \styp | W \mspace{-4mu} = \mspace{-4mu} 1 \mspace{-2mu} \right\} \\
	& \mspace{8mu} \le \mspace{-20mu} \sum_{w=2}^{2^{K(\tau_x + \tau_n)R}} \mspace{-25mu} 2^{-K(I([T_x]_{\mathcal{P}};[Y]_{\mathcal{Q}}) - \epsilon_1)} \\
	& \mspace{8mu} \le 2^{-K(I([T_x]_{\mathcal{P}};[Y]_{\mathcal{Q}}) - (\tau_x + \tau_n)R - \epsilon_1)},
\end{align*}

\vspace{-0.15cm}
\noindent which goes to zero as $K \to \infty$ if $R < \frac{I([T_x]_{\mathcal{P}};[Y]_{\mathcal{Q}}) - \epsilon_1}{\tau_x + \tau_n}$. Next, we note that $I(T_x;Y) = \sup_{\mathcal{P},\mathcal{Q}} I([T_x]_{\mathcal{P}};[Y]_{\mathcal{Q}})$, see \cite[eq. 8.54]{cover-book}. Thus, combining the bounds on $\Pr \{\mathcal{E}_l | W = 1\}, l=1,2$, we have that $R < \mathsf{C}(\tau_n)$ is achievable.

\vspace{-0.1cm}
\subsection{Converse}

\vspace{-0.05cm}
Let $P_e^{(K)} \to 0$ as $K \to \infty$, for a sequence of encoders and decoders $\varphi^{(K)}$ and $\nu^{(K)}$. By Fano's inequality \cite[Theorem 2.10.1]{cover-book} we have:
\vspace{-0.2cm}
\begin{align}
	H(W|\hat{W}) \le 1 + P_e^{(K)} K(\tau_x + \tau_n) R \le K \delta(P_e^{(K)}), \label{eq:Fano_1}
\end{align}

\vspace{-0.15cm}
\noindent where $\delta(x)$ is a non-negative function that approaches $\frac{1}{K}$ as $x \to 0$. Next, observe that:
\begin{align}
	H(W|\hat{W}) \mspace{-3mu} \stackrel{(a)}{\ge} \mspace{-3mu} H \mspace{-2mu} \left(W|\hat{W}, \{Y_{k}\}_{k=1}^K \right) \mspace{-3mu} \stackrel{(b)}{\ge} \mspace{-3mu} H \mspace{-2mu} \left(W| \{Y_{k}\}_{k=1}^K \right), \label{eq:Fano_2}
\end{align}

\noindent where (a) follows from the fact that conditioning reduces entropy, and (b) follows from the fact that $\hat{W}$ is a function of $\{Y_{k}\}_{k=1}^K$. We now write:
\begin{align}
	K R (\tau_x + \tau_n) & = H(W) \label{eq:converse_1} \\
	& = I(W; \{Y_{k}\}_{k=1}^K) + H \left( W| \{Y_{k}\}_{k=1}^K \right) \nonumber \\
	& \le I(W; \{Y_{k}\}_{k=1}^K) + K \delta(P_e^{(K)}) \label{eq:converse_3} \\
	& = \sum_{i=1}^{K} I(W; Y_i | \{Y_{k}\}_{k=1}^{i-1}) + K \delta(P_e^{(K)}) \nonumber \\
	& \le \sum_{i=1}^{K} I(W, \{Y_{k}\}_{k=1}^{i-1}; Y_i) + K \delta(P_e^{(K)}) \label{eq:converse_5} \\
	& = \sum_{i=1}^{K} I(T_{x,i}, W, \{Y_{k}\}_{k=1}^{i-1}; Y_i) + K \delta(P_e^{(K)}) \label{eq:converse_6} \\
	& = \sum_{i=1}^{K} I(T_{x,i}; Y_i) + K \delta(P_e^{(K)}) \label{eq:converse_7} \\
	& \le K \underset{\tau_x, \mathcal{F}(\tau_x)}{\max} I(T_x; Y) + K \delta(P_e^{(K)}), \nonumber
\end{align}

\noindent where \eqref{eq:converse_1} follows from the fact the the messages are uniformly distributed; \eqref{eq:converse_3} follows from Fano's inequality, see \eqref{eq:Fano_1}--\eqref{eq:Fano_2}; \eqref{eq:converse_5} follows from the non-negativity of mutual information; \eqref{eq:converse_6} follows from the fact that $T_{x,i}$ is a function of $W$, and \eqref{eq:converse_7} follows from the fact that the channel is memoryless. Thus, the above chain of inequalities implies that $R < \mathsf{C}(\tau_n) + \frac{\delta(P_e^{(K)})}{\tau_x + \tau_n}$, which tends to $\mathsf{C}(\tau_n)$ when $K \to \infty$. This completes the proof of the converse.

\section{Proof of Lemma \ref{lem:partEntLevy}}
\label{app:ProofPartEntLevy}

\begin{figure*}[t]
	\normalsize
	\setcounter{MYtempeqncnt}{\value{equation}}
	\setcounter{equation}{70}
	
	\begin{align}
	\frac{3\sqrt{c}}{2\sqrt{2\pi}}\int_0^\tau x^{-3/2}\exp\bigg( \frac{-c}{2t}\bigg)\log(x) dx & = \frac{3\sqrt{c}}{2\sqrt{2\pi}} \frac{\Gamma(\tfrac{1}{2},\tfrac{c}{2\tau}) \log(c/2)-\Gamma^\prime(\tfrac{1}{2},\tfrac{c}{2\tau})\log(e) }{\sqrt{\tfrac{c}{2}}} \nonumber \\
	& = \frac{3[\Gamma(\tfrac{1}{2},\tfrac{c}{2\tau}) \log(\tfrac{c}{2})-\log(\tfrac{c}{2\tau})\Gamma(\tfrac{1}{2},\tfrac{c}{2\tau})-\tfrac{c}{2\tau}T(3,\tfrac{1}{2},\tfrac{c}{2\tau})\log(e)]}{2\sqrt{\pi}} \label{eq:L3PI1US}
	\end{align}
	
	\hrulefill
	
	\setcounter{equation}{\value{MYtempeqncnt}}
	
\end{figure*}

	Let $X$ be a \levy distributed RV. Then, plugging \eqref{eqn:LevyPDF_0} into \eqref{eq:funcN} we write:
	\begin{align}
	\eta(X,\tau) &= -\int_0^\tau f_X(x) \bigg( \frac{1}{2}\log\bigg(\frac{c}{2\pi}\bigg) \noindent \\
	& \quad \qquad -\frac{3}{2}\log(x) -\frac{c}{2\ln(2)x} \bigg) dx \nonumber \\
	&=\frac{1}{2}\log\bigg(\frac{2 \pi}{c}\bigg)F_X(\tau) +\int_0^\tau f_X(x) \frac{3}{2}\log(x) dx \nonumber \\   
	& \qquad \qquad \qquad + \int_0^\tau f_X(x)\frac{c}{2\ln(2)x} dx, \label{eq:L3integralForm}
	\end{align}
	
	\noindent where $F_X(x)$ is given in \eqref{eqn:LevyCDF}.	
	To solve the integrals in \eqref{eq:L3integralForm} we introduce the following lemma:
	\begin{lemma}
		\label{lem:incompleteInt}
		Let $\Gamma(s,x), s,x > 0$, be the incomplete gamma function \cite{nist10} given by:
		\begin{align}
		\Gamma(s,x) = \int_x^\infty y^{s-1} e^{-y} dy, \label{eq:GammaDef}
		\end{align}
		
		\noindent and let $\Gamma^\prime(s,x)$ be its derivative with respect to the first parameter $s$, given by \cite[eq. (29)]{ged90}:
		\begin{align}
		\Gamma^\prime(s,x) = \ln(x)\Gamma(s,x)+xT(3,s,x), \label{eq:GammaDeriv}
		\end{align}
		
		\noindent where $T(3,s,x)$ is a special case of the Meijer G-function given in \cite[eq. (31)]{ged90}. 
		Then, the following holds for $m, a, n, \tau > 0$:
		\begin{align}
		\label{eq:Lem1int1}
		\int_0^\tau x^{-mn-1} &\exp\bigg(-\frac{a}{x^n}\bigg) dx = \frac{\Gamma(m,a/\tau)}{na^m}
		\end{align}
		\begin{align}
		\label{eq:Lem1int2}
		\int_0^\tau x^{-mn-1} &\exp\bigg(-\frac{a}{x^n}\bigg) \log(x) dx = \nonumber \\
		&\frac{\Gamma(m,a/\tau)\log(a)-\Gamma^\prime(m,a/\tau)\log(e)}{n^2a^m}
		\end{align}
	\end{lemma}
	
	\begin{IEEEproof}
	The proof of is provided in Appendix~\ref{app:ProofLemma1}.
	\end{IEEEproof}
	
	\setcounter{equation}{71}
	
	Now, the first integral in \eqref{eq:L3integralForm} can be solved using \eqref{eq:Lem1int2} as \eqref{eq:L3PI1US} at the top of the next page.
	
	\noindent Using {\tt Mathematica} we can write the function $T(3,\tfrac{1}{2},\tfrac{c}{2\tau})$ as:\footnote{The {\tt Mathematica} command is:
	\begin{equation*}
	\text{\tt MeijerG[\{\{\}, \{0, 0\}\}, \{\{-1, -1/2, -1\}, \{\}\}, $\frac{c}{2 \tau}$]}.
	\end{equation*}}	 
	\begin{align*}
	&T(3,\tfrac{1}{2},\tfrac{c}{2\tau}) =  \nonumber\\
	&\frac{\tau\bigg[ 4 \sqrt{\tfrac{c}{2\tau}}~_2F_2(\tfrac{1}{2},\tfrac{1}{2},\tfrac{3}{2},\tfrac{3}{2};\tfrac{-c}{2\tau}) \mspace{-3mu} - \mspace{-3mu} \sqrt{\pi}\ln(\tfrac{c}{2\tau}) \mspace{-3mu} - \mspace{-3mu} \sqrt{\pi}(\gamma \mspace{-3mu} + \mspace{-3mu} \ln4)\bigg]}{c/2},
	\end{align*}
	
	\noindent where $\gamma$ is the Euler's constant and $_pF_q(\cdot)$ is the generalized hypergeometric function \cite[Ch. 16]{nist10}. Let $g(c,\tau)$ $=~_2F_2(\tfrac{1}{2},\tfrac{1}{2},\tfrac{3}{2},\tfrac{3}{2};\tfrac{-c}{2\tau})$. Using the property \cite[eq. (8.4.14)]{nist10}:
	\begin{align}
	\label{eq:InGam0.5}
	\Gamma (\tfrac{1}{2},x^2) = \sqrt{\pi} \erfc(x),
	\end{align}
	
	\noindent \eqref{eq:L3PI1US} can be simplified to:
	\begin{align}
	& \frac{3\sqrt{c}}{2\sqrt{2\pi}}\int_0^\tau x^{-3/2}\exp\bigg( \frac{-c}{2t}\bigg)\log(x) dx \nonumber \\
	& \quad =\frac{3}{2}\bigg((\erfc(\sqrt{\tfrac{c}{2\tau}})-1) \log(\tau)-4\sqrt{\tfrac{c}{2\pi\tau}}g(c,\tau)\log(e) \nonumber\\
	& \mspace{150mu}+\log(c/2)+\gamma\log(e)+2\bigg). \label{eq:L3FirtIntTerm}
	\end{align}
	
	\noindent The second integral in (\ref{eq:L3integralForm}) can be solved using (\ref{eq:Lem1int1}) as:
	\begin{align*}
	\frac{c^{3/2}}{2\ln(2)\sqrt{2\pi}} \int_0^\tau x^{-5/2}\exp(-\tfrac{c}{2x})dx=\frac{\Gamma(\tfrac{3}{2},\tfrac{c}{2\tau})}{\ln(2)\sqrt{\pi}}.
	\end{align*}
	
	\noindent Using \eqref{eq:InGam0.5} and the property \cite[eq. (8.8.2)]{nist10}:
	\begin{align*}
	\Gamma(s+1,x) = s \Gamma(s,x) + x^se^{-x},
	\end{align*}
	
	\noindent we obtain:
	\begin{align}
	\label{eq:L3SecIntForm}
	\int_0^\tau \mspace{-4mu} \frac{c f_X(x)}{2\ln(2)x} dx \mspace{-4mu} = \mspace{-4mu} \log(e)\bigg( \mspace{-2mu} \tfrac{1}{2}\erfc(\sqrt{\tfrac{c}{2\tau}}) \mspace{-3mu} + \mspace{-3mu} \sqrt{\tfrac{c}{2\pi\tau}}e^{-\tfrac{c}{2\tau}} \mspace{-2mu} \bigg).
	\end{align}
	
	\noindent Finally, plugging \eqref{eqn:LevyCDF}, \eqref{eq:L3FirtIntTerm}, and \eqref{eq:L3SecIntForm} into \eqref{eq:L3integralForm} concludes the proof.

\section{Proof of Lemma \ref{lem:incompleteInt}}
\label{app:ProofLemma1}
	First, we consider \eqref{eq:Lem1int1} and write:
	\begin{align*}
	& \int_0^\tau x^{-mn-1} \exp\bigg(\frac{-a}{x^n}\bigg) dx \\
	&  \qquad = \frac{-1}{a^mn} \int_0^\tau \bigg(\frac{a}{x^n} \bigg)^{m-1} \mspace{-5mu} \exp\bigg(\frac{-a}{x^n} \bigg) \bigg(\frac{-an}{x^{n+1}} \bigg) dx.		
	\end{align*}
	
	\noindent Substituting $y=a/x^n$, and $dy = -an/x^{-n+1} dx$ we obtain:
	\begin{align}
	&\frac{-1}{a^mn} \int_0^\tau \bigg(\frac{a}{x^n} \bigg)^{m-1} \mspace{-5mu} \exp\bigg(\frac{-a}{x^n} \bigg) \bigg(\frac{-an}{x^{n+1}} \bigg) dx \nonumber \\
	& \qquad \quad = \frac{-1}{a^mn} \int_\infty^{a/\tau} y^{m-1}\exp(-y) dy \nonumber \\
	& \qquad \quad = \frac{1}{a^mn} \int_{a/\tau}^\infty y^{m-1}\exp(-y) dy \nonumber \\
	& \qquad \quad = \frac{\Gamma(m,a/\tau )}{n a^m}, \label{eq:gammaNormalized}
	\end{align}
	
	\noindent where \eqref{eq:gammaNormalized} follows from the definition of $\Gamma(\cdot, \cdot)$ in \eqref{eq:GammaDef}.
	
	To prove \eqref{eq:Lem1int2}, we write:
	\begin{align*}
	& \frac{\partial }{\partial m} \left( \int_0^\tau x^{-mn-1} \exp\bigg(\frac{-a}{x^n}\bigg) dx \right) \\
	& \qquad = \frac{-n}{\log e} \int_0^\tau \log(x) x^{-mn-1} \exp\bigg(-\frac{a}{x^n}\bigg) dx.
	\end{align*}
	
	\noindent Clearly, this equals to the derivative of \eqref{eq:gammaNormalized}. Thus, we obtain:
	\begin{align*}
	\frac{-n}{\log e} \int_0^\tau \log(x) x^{-mn-1} \exp\bigg(\frac{-a}{x^n}\bigg) dx = \frac{\partial}{\partial m} \bigg[\frac{\Gamma(m,a/\tau )}{a^m n} \bigg].
	\end{align*}
	
	\noindent By using \eqref{eq:GammaDeriv} and organizing the terms we obtain \eqref{eq:Lem1int2}. 

\section{Proof of Corollary \ref{cor:explicitUB_value}} \label{annex:cor_explicitUB_value_proof}

First, we recall the maximization problem in \eqref{eq:capacityUB_single}, which can be stated as follows:
\begin{align}
	 \max_{x \ge 0} g(x) \triangleq \frac{\left(\log(x + \alpha) \mspace{-3mu} - \mspace{-3mu} \beta \right)\gamma}{x+\alpha}, \label{eq:optProb_ub}
\end{align}

\noindent for $0 < \alpha, 0 \le \gamma \le 1$, and $\beta < \log(\alpha)$. The last constraint follows from the fact that $h(T_n | T_n \le \tau_n)$ is the entropy of an RV with the support of size $\tau_n$. 
Next, note that the derivative of $g(x)$ is givn by:
\begin{align*}
	g'(x) = \frac{(1 + \beta - \log(x + \alpha))\gamma}{(x + \alpha)^2}.
\end{align*}

\noindent Thus, the extrema of $g(x)$ is given by:
\begin{align}
	x^{\ast} = 2^{(1 + \beta)} - \alpha.
\end{align}

\noindent Now, since $g(0) > 0$, and $\lim_{x \to \infty} g(x) = 0$, we conclude that $x^{\ast}$ is a maxima. Thus, the maximizing $x$ over the range $x \ge 0$ is given by $x^{\ast} = \max \{0, 2^{(1 + \beta)} - \alpha\}$. Plugging these two values into \eqref{eq:optProb_ub} concludes the proof of \eqref{eq:capacityUB_single}.

To find the maximizing $\tau_x$ for \eqref{eq:capacityLB_single} we follow the same lines and note that:
\begin{align}
	& \mspace{-10mu} \frac{\partial}{\partial \tau_x} \frac{\left(m(\tau_x,\tau_n,T_n) \mspace{-3mu} - \mspace{-3mu} h(T_n|T_n\leq\tau_n)\right)F_{T_n}(\tau_n)}{\tau_x + \tau_n} \nonumber \\
	& \mspace{10mu} = \frac{\tau_x}{(\tau_x + \tau_n)(\tau_x^2 + 4^{h(T_n|T_n\leq\tau_n)})} \nonumber \\
	& \mspace{40mu} - \frac{0.5 \log(\tau_x^2 + 4^{h(T_n|T_n\leq\tau_n)}) - h(T_n|T_n\leq\tau_n)}{(\tau_x + \tau_n)^2}.
\end{align}

\noindent Thus, equating this derivative to zero we obtain the required equation.

\section{Proof of Lemma \ref{lm:MinLevyGumbel}}
\label{app:ProofMinLevyGumbel}

Recall that $F_{T_n}^{-1}(\cdot)$ is the inverse CDF of a L\'evy-distributed RV, given by:
\begin{align*}
F_{T_n}^{-1}(u) = \frac{c}{2\erfcinv^2(u)}.
\end{align*}

\noindent Further note that 
\begin{align*}
	\lim_{\epsilon\rightarrow\infty} \frac{F_{T_n}^{-1}(\epsilon)-F_{T_n}^{-1}(2\epsilon)}{F_{T_n}^{-1}(2\epsilon)-F_{T_n}^{-1}(4\epsilon)} =1.
\end{align*}

\noindent Therefore, \cite[Theorem 3.9]{ExtremeValueBook} implies that for sufficiently large $M$, $\tilde{T}_n$, the minimum of $M$ i.i.d. L\'evy-distributed RVs, belongs to Gumbel type domain of attraction. Moreover, using \cite[Theorem 3.2 and Theorem 3.4]{ExtremeValueBook} we obtain the expressions for the parameters of this limiting Gumbel distribution, see $\alpha$ and $\beta$ in \eqref{eq:paramGumb}.

%

\section{Proof of Lemma \ref{lm:condGumbEntropy}}
\label{app:ProofPartEntGumbel}

Let $\tilde{Z} \sim\GumbDist(\alpha,\beta)$. Then, plugging \eqref{eq:GumbPDF} into \eqref{eq:funcN} we write:
\begin{align}
\eta(\tilde{Z},\tau) &= -\int_{-\infty}^\tau f_Z(z) \bigg[\log\bigg(\frac{1}{\beta}\bigg) \noindent \\
& \quad \qquad +\log\bigg(\exp\left[\tfrac{z-\alpha}{\beta}-\exp\left(\tfrac{z-\alpha}{\beta}\right) \right]\bigg)  \bigg] dz \nonumber \\
&=F_Z(\tau)\log(\beta) \nonumber \\   
& \qquad  - \log(e)\int_{-\infty}^\tau f_Z(z)\left[\tfrac{z-\alpha}{\beta}-\exp\left(\tfrac{z-\alpha}{\beta}\right) \right] dz, \label{eq:GumbIntegralProof1}
\end{align}
where $F_Z(z)$ is given in \eqref{eq:GumbCDF}. Using the change of variable $u=\tfrac{z-\alpha}{\beta}$, $a=\tfrac{\tau-\alpha}{\beta}$, and $du=\tfrac{dz}{\beta}$, the integral in \eqref{eq:GumbIntegralProof1} can be written as
\begin{align}
	& \mspace{-8mu} - \log(e)\int_{-\infty}^a \exp\left[u-\exp\left(u\right) \right] \left[u-\exp\left(u\right) \right] du = \nonumber \\
    & \mspace{35mu} \log(e)\left[(a \mspace{-3mu} - \mspace{-3mu} 1)e^{-e^a} \mspace{-3mu} + \mspace{-3mu} e^{a-e^a} \mspace{-3mu} + \mspace{-3mu} 1 \mspace{-3mu} + \mspace{-3mu} \gamma \mspace{-3mu} - \mspace{-3mu} \Ei(-e^a) \right],
    \label{eq:GumbIntegralProof2}
\end{align}
	where $\gamma \mspace{-3mu} \approx \mspace{-3mu} 0.5772$ is the Euler's constant \cite[Ch. 5.2]{nist10}, and $\Ei(\cdot)$ is the exponential integral \cite[Equation 6.2.5]{nist10}. Substituting $a=\tfrac{\tau-\alpha}{\beta}$ into \eqref{eq:GumbIntegralProof2}, and \eqref{eq:GumbIntegralProof2} into \eqref{eq:GumbIntegralProof1}, we obtain \eqref{eq:condiEntGumb}.

\section{Scaling of the Numerator of \eqref{eq:capacityLB_FA}}
\label{app:ProofScallingFA}

Recall that $\tilde{T}_{n,k} = \min (\vec{T}_{n,k})$, and note that the PDF of $\tilde{T}_{n,k}$ concentrates towards zero with increasing $M$, which leads to the Gumbel domain of attraction. Thus, 
\begin{align}
	\Pr \{ \tilde{T}_{n,k} \le \tau_n \} \to_{M \to \infty} 1. \label{eq:FA_indep_of_particleLifeTime}
\end{align}

\noindent We are interested in deriving the scaling of $m(\tau_x,\tau_n,\tilde{T}_n) \mspace{-3mu} - \mspace{-3mu} h(\tilde{T}_n|\tilde{T}_n\leq\tau_n)$ as $M \mspace{-3mu} \to \mspace{-3mu} \infty$. 
%
%
\noindent For this purpose, for sufficiently large $M$, we write \eqref{eqn:mFuncDef} as:
\begin{align}
		m(\tau_x,\tau_n,\tilde{T}_{n}) & = 0.5 \log \big( \tau_x^2+2^{2h(\tilde{T}_n|\tilde{T}_n\leq\tau_n)}\big) \nonumber \\
		& \approx 0.5 \log \big( \tau_x^2+2^{2h(\tilde{T}_n)}\big), \label{eq:FA_mFunc} \\
		& \approx \log \tau_x, \label{eq:FA_limiting_mFunc}
	\end{align} 
	
	\noindent where \eqref{eq:FA_mFunc} follows from \eqref{eq:FA_indep_of_particleLifeTime}, and \eqref{eq:FA_limiting_mFunc} follows from the fact that the density of $\tilde{T}_n$ concentrates towards zero, and therefore $h(\tilde{T}_n) \to -\infty$ when $M \to \infty$.
	Therefore, for sufficiently large $M$, $m(\tau_x,\tau_n,\tilde{T}_n) \mspace{-3mu} - \mspace{-3mu} h(\tilde{T}_n|\tilde{T}_n\leq\tau_n)$ can be approximated by:
	\begin{align}
		m(\tau_x,\tau_n,\tilde{T}_n) \mspace{-3mu} - \mspace{-3mu} h(\tilde{T}_n|\tilde{T}_n\leq\tau_n) \approx \log \tau_x \mspace{-3mu} - \mspace{-3mu} h(\tilde{T}_n).
	\end{align}
	
	We now evaluate the scaling of $h(\tilde{T}_n)$. Recall that for sufficiently large $M$, $\tilde{T}_n \sim \GumbDist(\alpha, \beta)$ with $\alpha$ and $\beta$ given in \eqref{eq:paramGumb}. Thus, the entropy of $\tilde{T}_n$ is given by:
	\begin{align}
		h(\tilde{T}_n) \approx \log_e (\beta) + \gamma + 1,
	\end{align}
	
	\noindent where $\gamma \mspace{-3mu} \approx \mspace{-3mu} 0.5772$ is the Euler's constant \cite[Ch. 5.2]{nist10}. 
	Hence, to obtain the scaling of $h(\tilde{T}_n)$ we evaluate the scaling of $\beta$, given in \eqref{eq:paramGumb} by:
	\begin{align}
	\beta = \frac{c}{2} \left( \frac{1}{\erfcinv^2(\tfrac{1}{M})} - \frac{1}{\erfcinv^2(\tfrac{1}{Me}) } \right). \label{eq:Gumb_beta}
	\end{align}
	
	To approximate $\beta$ we first note that $\erfcinv(x) = \erfinv(1-x), 0\le x \le 1$, and approximate $\erfinv(x)$, for large $x$, using \cite[eq. (13)]{Strecok68}:
	\begin{align}
		\erfinv(x) \approx \sqrt{- \log_e \left(1 - x^2 \right) }.	\label{eq:erfinv_apprx}
	\end{align}
	
	\noindent Explicitly, $\beta$ can be approximated as:
	\begin{align}
		\beta & = \frac{c}{2} \left( \frac{1}{\erfinv^2(1 - \tfrac{1}{M})} - \frac{1}{\erfinv^2(1 -\tfrac{1}{Me}) } \right) \nonumber \\
		& \approx \frac{c}{2} \left( \frac{1}{\log_e (1 - (1 - \tfrac{1}{Me})^2)} - \frac{1}{ \log_e (1 - (1 - \tfrac{1}{M})^2)} \right) \nonumber \\
		& = \frac{c}{2} \left( \frac{1}{\log_e (\tfrac{2}{Me} - \tfrac{1}{M^2 e^2}))} - \frac{1}{ \log_e (\tfrac{2}{M} - \tfrac{1}{M^2})} \right) \nonumber \\
		& \approx \frac{c}{2} \left( \frac{1}{\log_e (\tfrac{2}{Me})} - \frac{1}{ \log_e (\tfrac{2}{M})} \right) \nonumber \\
		& = \frac{c}{2} \frac{\log_e (\tfrac{2}{M}) - \log_e (\tfrac{2}{Me})}{ \log_e (\tfrac{2}{Me}) \log_e (\tfrac{2}{M})} \nonumber \\
		& = \frac{c}{2} \frac{1}{ \log_e (\tfrac{2}{Me}) \log_e (\tfrac{2}{M})} \nonumber \\
		& = \frac{c}{2} \frac{1}{ \log_e^2 (\tfrac{2}{M}) - \log_e (\tfrac{2}{M})} \nonumber \\
		& \approx \frac{c}{2} \frac{1}{ \log_e^2 (\tfrac{2}{M}) }.
	\end{align}
	
	\noindent Therefore, for sufficiently large $M$, we obtain:
	\begin{align}
		h(\tilde{T}_n) & \approx \log_e (\beta) \nonumber \\
		& \approx \log_e \left( \frac{c}{2} \frac{1}{ \log_e^2 (\tfrac{2}{M})} \right) \nonumber \\
		& = \log_e \frac{c}{2} - \log_e \left( \log_e^2 (\tfrac{2}{M}) \right) \nonumber \\
		& = \log_e \frac{c}{2} - 2 \log_e \left( \log_e (\tfrac{M}{2}) \right),
	\end{align}
	
	\noindent which leads to the desired scaling.

\section{Proof of Lemma \ref{lm:convToGausChan}}
\label{app:ProofLemmaConvGausChan}	

Since the channel is memoryless, in the following we drop the subscript $k$. As $M\rightarrow\infty$, $\mathbb{P}(|\mathcal{J}| = 0)\rightarrow0$, and we can write:
\begin{align}
	\label{eq:app:equivalenAvgChan}
	\tfrac{1}{|\mathcal{J}|}\underset{i\in\mathcal{J}}{\sum}(\vec{T}_{y}[i] \mspace{-3mu} -\mathbb{E}[T^\prime_{n}]) = \mspace{-3mu} T_{x} \mspace{-3mu} + \mspace{-3mu} \tfrac{1}{|\mathcal{J}|}\underset{i\in\mathcal{J}}{\sum}(\vec{T}^\prime_{n}[i]-\mathbb{E}[T^\prime_{n}]),
\end{align}	

\noindent where $\vec{T}^\prime_{n}$ is a vector of i.i.d. truncated \levy RVs with parameters $c$ and $\tau_n$, and $\mathbb{E}[T^\prime_{n}]$ is the mean of a truncated \levy RV with parameters $c$ and $\tau_n$. Note that we can assume that the receiver subtracts this mean value from each arrival time, since this parameter is known at the receiver. Therefore, equivalently, this channel can be written using i.i.d. truncated \levy RVs with zero means. Let $\vec{T}\dprime_{n}[i]=\vec{T}^\prime_{n}[i]-\mathbb{E}[T^\prime_{n}]$ represent this zero mean truncated \levy noise, and $\vec{T}\dprime_{y}[i]=\vec{T}_{y}[i]-\mathbb{E}[T^\prime_{n}]$ the corresponding channel output. Then the channel output is given by:
\begin{align}
\label{eq:app:equivalenAvgChan2}
\tfrac{1}{|\mathcal{J}|}\underset{i\in\mathcal{J}}{\sum}\vec{T}\dprime_{y}[i] = \mspace{-3mu} T_{x} \mspace{-3mu} + \mspace{-3mu} \tfrac{1}{|\mathcal{J}|}\underset{i\in\mathcal{J}}{\sum}\vec{T}\dprime_{n}[i].
\end{align}

Let $\mathds{1}_i$ be the indicator that the \ith particle arrives at the receiver, i.e., $i\in\mathcal{J}$. Since the particles arrive independently with probability $F_{T_n}(\tau_n)$, this indicator function is characterized by this probability. Furthe, let $\vec{Z}_n[i]=\mathds{1}_i\times\vec{T}\dprime_{n}[i]$, where $\var[Z_n]=F_{T_n}(\tau_n)\var[T^\prime_{n}]$, and finally let $W\sim\NormDist\left(0,\tfrac{\var[Z_n]}{M}\right)$. Then, the channel in \eqref{eq:app:equivalenAvgChan2}, as $M\rightarrow\infty$, can be written as
\begin{align}
\tfrac{1}{|\mathcal{J}|}\underset{i\in\mathcal{J}}{\sum}\vec{T}\dprime_{y}[i] &= \mspace{-3mu} T_{x} \mspace{-3mu} + \mspace{-3mu} \tfrac{1}{M F_{T_n}(\tau_n)}\tfrac{M F_{T_n}(\tau_n)}{|\mathcal{J}|}\sum_{i=1}^M \vec{Z}\dprime_{n}[i] \label{eq:app:equivalenAvgChan3.1}\\
&=\mspace{-3mu} T_{x} \mspace{-3mu} + \mspace{-3mu} \tfrac{1}{F_{T_n}(\tau_n)}\tfrac{M F_{T_n}(\tau_n)}{|\mathcal{J}|}W \label{eq:app:equivalenAvgChan3.2}\\
&=\mspace{-3mu} T_{x} \mspace{-3mu} + \mspace{-3mu} \tfrac{1}{F_{T_n}(\tau_n)}W \label{eq:app:equivalenAvgChan3.3}\\
&=\mspace{-3mu} T_{x} \mspace{-3mu} + \hat{T}_n\mspace{-3mu}, \nonumber
\end{align}	
where \eqref{eq:app:equivalenAvgChan3.2} follows due to the central limit theorem and \eqref{eq:app:equivalenAvgChan3.3} follows by law of large numbers.

\section{The Entropy of \levy Distributed RVs}
\label{app:ProofEntLevy}

To derive the entropy expression of a L\'evy distributed RV, we consider Lemma \ref{lem:incompleteInt} while setting the integral upper boundary to be $\infty$, i.e., $\tau \to \infty$. The resulting integrals are presented in the following lemma.
\begin{lemma}
	The following two integral equations hold:
	\begin{align}
	\label{eq:Lem2int1}
	\int_0^\infty x^{-mn-1} &\exp\bigg(-\frac{a}{x^n}\bigg) dx = \frac{\Gamma(m)}{na^m},
	\end{align}
	\begin{align}
	\label{eq:Lem2int2}
	\int_0^\infty \log(x) x^{-mn-1} &\exp\bigg(-\frac{a}{x^n}\bigg) dx = \nonumber \\
	&\frac{\Gamma(m)\log(a)-\Gamma^\prime(m)\log(e)}{n^2a^m},
	\end{align}
	for $m>0$, $a>0$, and $n>0$, where  $\Gamma(\cdot)$ is the gamma function \cite[eq. (5.2.1)]{nist10} given by:
	\begin{align}
	\Gamma(s) = \int_0^\infty y^{s-1} e^{-y} dy;
	\end{align}
	and $\Gamma^\prime(\cdot)$ is its derivative with respect to the parameter $s$.
\end{lemma}
\begin{IEEEproof}
	The proof follows steps similar to the steps taken in the proof of Lemma \ref{lem:incompleteInt}, and hence, it is omitted.
\end{IEEEproof}

Next, we recall that entropy is invariant to time shifts and therefore we assume that $\mu = 0$ and write:
\begin{align}
	h(X) &= -\int_0^\infty f_X(x) \log (f_X(x)) dx \nonumber \\
	& = \mspace{-3mu} - \mspace{-2mu} \int_0^\infty \mspace{-9mu} f_X(x) \mspace{-2mu} \left( \mspace{-2mu} \frac{1}{2}\log\bigg( \mspace{-2mu} \frac{c}{2\pi} \mspace{-2mu} \bigg) \mspace{-3mu} - \mspace{-3mu}\frac{3}{2}\log(x) \mspace{-3mu} - \mspace{-3mu} \frac{c}{2\ln(2)x} \mspace{-2mu} \right) \mspace{-4mu} dx \nonumber \\
	&=\frac{1}{2}\log\bigg(\frac{2 \pi}{c}\bigg) +\int_0^\infty f_X(x) \frac{3}{2}\log(x) dx \nonumber \\   
  & \mspace{150mu} + \int_0^\infty f_X(x)\frac{c}{2\ln(2)x} dx. \label{eq:T1integralForm}
\end{align}

\noindent Using \eqref{eq:Lem2int2}, we write the first integral in \eqref{eq:T1integralForm} as:
\begin{align}
	& \frac{3\sqrt{c}}{2\sqrt{2\pi}}\int_0^\infty x^{-3/2}\exp\bigg( \frac{-c/2}{t}\bigg)\log(x) dx  \nonumber \\
	& \qquad \qquad = \frac{3\sqrt{c}}{2\sqrt{2\pi}} \frac{\Gamma(1/2) \log(c/2)-\Gamma^\prime(1/2)\log(e) }{\sqrt{c/2}} \nonumber \\
	& \qquad \qquad = \frac{3\Gamma(1/2)[\log(c/2)-\psi(1/2)\log(e)]}{2\sqrt{\pi}} \nonumber \\
	& \qquad \qquad = \frac{3\sqrt{\pi}[\log(c/2)-(-\gamma-2\ln(2))\log(e)]}{2\sqrt{\pi}} \nonumber \\
	& \qquad \qquad = \frac{3[\log(c/2)+\gamma\log(e)+2]}{2}, \label{eq:T1FirstIntTerm}
\end{align}

\noindent where $\psi(s)=\Gamma^\prime(s)/\Gamma(s)$ is the digamma or Psi function, $\Gamma(1/2)=\sqrt{\pi}$, and $\psi(1/2)= - \gamma-2\ln2$ as shown in\cite[Ch. 5.4]{nist10}. 
Similarly, using (\ref{eq:Lem2int1}) and the fact that $\Gamma(3/2)=0.5\sqrt{\pi}$, it can be shown that the second integral reduces to $\log(e)/2$. Substituting these solutions into (\ref{eq:T1integralForm}) and simplifying we conclude the proof.

\bibliographystyle{IEEEtran}
\bibliography{IEEEabrv,MolCom}

\end{document}